\documentclass[12pt]{article}
\usepackage{amsmath}
\usepackage{bm}
\usepackage{bbm}
\usepackage{cite}
\usepackage[dvips]{color}

\usepackage{graphicx}
\usepackage{slashed}
\usepackage{tabularx}
\usepackage{comment}
\usepackage{hyperref} 

\newcommand{\Sec}[1]{section~\ref{#1}}

\begin{document}
\title{
\vspace{-1.2truecm}
\Large\bf
 Spectra of cosmic ray electrons and diffuse gamma rays with the constraints of AMS-02 and HESS data 
} 
\author{
  Ding Chen$^a$, Jing Huang$^b$ and Hong-Bo Jin$^a$\footnote{Corresponded author Email: hbjin@bao.ac.cn}\\
  \small\textit{$^a$National Astronomical Observatories, Chinese Academy of Sciences, Beijing, 100012, China}\\ 
  \small\textit{$^b$Institute of High Energy Physics, Chinese Academy of Sciences, Beijing 100049, China}\\
}
\date{}
\maketitle
\begin{abstract}
Recently, AMS-02 reported their observed results of cosmic rays(CRs). In addition to the AMS-02 data, 
we add HESS data to estimate the spectra of CR electrons and the diffuse gamma rays above TeV.
In the conventional diffusion model, a global analysis is performed on the spectral features of CR electrons and the diffuse gamma rays by GALRPOP package. 
The results show that the spectrum structure of the primary component of CR electrons can not be fully reproduced by a simple power law and the relevant break is around hundred GeV. At 99\% C.L.,
the injection indices above the break decrease from 2.54 to 2.35, but the ones below the break are only in the range 2.746 - 2.751. 
The spectrum of CR electrons does not need to add TeV cutoff to match the features of HESS data too.
Based on the difference between the fluxes of CR electrons and the primary component of them, the predicted excess of CR positrons is consistent with the interpretations as pulsar or dark matter.
In the analysis of the Galactic diffuse gamma rays with the indirect constraint of AMS-02 and HESS data, it is found that the fluxes of Galactic diffuse gamma rays are consistent with GeV data of Fermi-LAT in the high latitude regions. The results indicate that the inverse Compton scattering(IC) is the dominant component in the range of the hundred GeV to tens of TeV respectively from the high latitude regions to the low ones, 
and in the all regions of Galaxy the flux of diffuse gamma rays is less than CR electrons at the energy scale of 20 TeV.  
\end{abstract}
\newpage
\section{Introduction}
Recently, AMS-02 reported their observed results of cosmic rays(CRs). Below TeV, the spectra of CR protons and electrons can be described by the high precision data\cite{Aguilar:2015ooa,Aguilar:2014mma}. 
However, above TeV, CR sepectra have many uncertainties. 
In the conventional model, CR production and propagation are governed by the same mechanism at energies below $10^{17}$ eV\cite{Ginzburg:1990sk}.
Thus, the TeV spectra of CRs may be predicted to match the data below TeV. 
In the pilot study of CR measurement, the predicted spectra of CRs are often used to analyze the background subtractions, identification of chemical composition, etc.
 
In the conventional model of CRs, CR electrons are divided into the primary and secondary particles.  As well as for the primary CR nucleons, the primary electrons are supposed to be created from Supernova remnants(SNRs) and the injection spectra can be described by a simple power law feature derived from the diffuse shock acceleration(DSA)\cite{Blandford:1987pw}. 
The secondary electrons and positrons are created during collisions of CR nucleons (protons dominant) with the interstellar gas, and also have a simple feature of spectra derived from the CR protons. The secondary electrons and positrons both contribute to the astrophysical background of CR electrons, as well as the primary electrons.  

In the recent years, it has been found that the ratio of the CR positron flux to the combined flux of CR electrons and positrons (positron fraction) keeps increasing in some energy ranges,
which is not consistent with the conventional astrophysical background data and called positron excess\cite{Adriani:2008zr,Adriani:2010ib,FermiLAT:2011ab,Aguilar:2013qda,Accardo:2014lma}.
The possible sources of positron excess are not only astrophysical origins, such as nearby pulsars\cite{Hooper:2008kg,Profumo:2008ms} and supernovae remnants(SNRs)\cite{Blasi:2009hv,Ahlers:2009ae} but also dark matter, which produces the excessive CR positrons through the annihilation or decay.
Lately, AMS-02 have reported the precise measurements of CR electrons and positrons\cite{Aguilar:2014mma} and updated the positron fraction\cite{Accardo:2014lma}. 
The flux of CR electrons is in the range of 0.5 to 700 GeV and cannot be fully described a single power-law spectrum\cite{Aguilar:2014mma}. The latest data of AMS-02 shows the features of positron excess explicitly.  

The spectral features of CRs are often used to explore their origins. In the AMS-02 data, as the maximal value of positron fraction is 0.159 at 305 GeV\cite{Accardo:2014lma}, the maximal flux of CR positrons is almost 20\% of the primary electrons. And the flux of CR positrons in the astrophysical background is about 1\% of the primary electrons at 305 GeV. 
Hence, if pulsars, dark matter or the other sources could produce the same flux of CR electrons as CR positrons, the experimental data of CR electrons would imply the flux of positron excess and has the special feature distinguished from a single power-law spectrum. 
In this paper, using the difference between the CR electrons and positrons fitting to AMS-02 data, we attempt to extract the astrophysical background of primary electrons to analyze the origins of CR electrons. 

As in the AMS-02 data, the maximum energy of CR positrons and electrons is below 1 TeV, in order to analyze the TeV flux of CR electrons, HESS experimental data are added. The measurements of HESS electrons are taken from an array of imaging atmospheric Cherenkov telescopes and do not discriminate the CR electrons from CR positrons (HESS electrons i.e. HESS electron and positron). 
The uncertainties of HESS electrons data are mainly from the subtraction of hadronic background and discrimination against gamma rays events\cite{Aharonian:2008aa}. The very-high-energy flux of HESS electrons is described by an exponentially cutoff power law with an index of 3.05$\pm$ 0.02 and a cutoff at 2.1$\pm$ 0.3 TeV in the range of 700 GeV to 5 TeV\cite{Aharonian:2008aa}. 
The low-energy extension of the HESS electron measurement  are from 340 GeV to 1.7 TeV with a break energy at about 1 TeV\cite{Aharonian:2009ah}. After HESS data is reported, the spectrum of CR electrons is often described by a broken power law with a break at 2 TeV\cite{Ackermann:2010ij}. 

Galactic diffuse gamma rays are related to with CRs interacting with the interstellar medium (ISM), which includes interstellar gas, interstellar radiation field(ISRF), magnetic field, etc. 
With the interstellar gas, CR nucleons produces neutral pions($\pi^0$), which decay into gamma rays\cite{dermer1986secondary}. 
As CR protons are dominant in the components of CR nucleons, the flux of $\pi^0$-decay diffuse gamma rays is mainly associated with CR protons. 
As above 10GV (rigidity of CRs) the measurement flux of CRs at earth is the same as the interstellar spectrum and below 10GV the interstellar spectrum may be transformed  into the observed value at earth by the solar modulation potential $\phi$\cite{Gleeson:1968zza}, the spectral features of diffuse gamma rays may be analyzed on the measurement data of CRs.
The experimental data of CR protons has been recently reported by CREAM\cite{Yoon:2011aa}, PAMELA\cite{Adriani:2011cu} and AMS-02\cite{Aguilar:2015ooa} experiments. Based on these measurement data, the flux of $\pi^0$-decay diffuse gamma rays can be calculated in the CR propagation model.
Besides the interaction of CR nucleons with ISM, CR electrons produce gamma rays by bremsstrahlung and by IC scattering with ISRF.  
Thus, the experimental data of diffuse gamma rays may be used to constraint indirectly the interstellar spectrum of CR electrons. 
Furthermore, with the constraint of CR particles and diffuse gamma rays data, the component discrimination of diffuse gamma rays can be performed to explore the distribution of ISM and ISRF. 

In this paper, we choose the diffuse gamma rays data of Milagro to constrain the IC component predicted by CR electrons above TeV.
The Milagro experiment is a water Cherenkov detector on air-shower arrays and reported the  the diffuse gamma ray spectra of inner Galaxy ($l\in[30^\circ,65^\circ]$) and Cygnus region($l\in[65^\circ,85^\circ]$) in the range of Galaxy latitude $b\in[-2^\circ,2^\circ]$\cite{Abdo:2008if}. 
The fluxes of inner Galaxy  are consisitent with prediction of the optimized GALPROP model \cite{Strong:2004ry,Prodanovic:2006bq,Porter:2008ve} and the fluxes of Cygnus region  exceed apparently the theoretical predictions of the background\cite{Abdo:2008if}.
As the subtractions of the isotropic background and the unknown point sources from the two regions are not indicated\cite{Abdo:2008if} and the predicted CR electrons of optimized GALPROP model are not verified by the experimental data yet,
the Milagro experimental data can provide an up-limit constraint to the spectrum of CR electrons predicted by AMS-02 and HESS data.

In this paper, based on the fluxes of CR electrons and protons,  all components of diffuse gamma rays are predicted in the conventional diffuse model by GALPROP package\cite{Strong:1998pw,Strong:2004ry,Prodanovic:2006bq,Porter:2008ve}.
In order to verify the predicted diffuse gamma rays below TeV, we choose the two regions of high latitudes in the public data of Fermi-LAT. 
As the fluxes of detected sources at those regions are weaker than the other regions\cite{FermiLAT:2012aa}, Fermi-LAT data may strongly constrain the prediction of the models based on AMS-02 and HESS data. 
We also analyze the components of diffuse gamma rays  and try to indicate some clues of gamma ray excess of Cygnus region. In details, the comparisons are also drawn between the fluxes of diffuse gamma rays, CR electrons and CR protons. 

This paper is organized as  follows. In \Sec{sec:CRpropagation}, we outline the formulas concerned with  the propagation of CRs and the calculation of diffuse gamma ray emission.
In \Sec{sec:fitSchemes}, we describe the data selection and the strategy of the data fitting in a number of propagation models.
In \Sec{sec:results}, we describe the analysis of CR electrons and diffuse gamma rays. 
The conclusions are given in \Sec{sec:conclusion}.

\section{Cosmic ray propagation and diffuse gamma ray emission}\label{sec:CRpropagation}
In the conventional model, CR production and propagation are governed by the same mechanism at energies below $10^{17}$ eV. CR propagation is often described by the diffusion equation\cite{Ginzburg:1990sk}:
\begin{align}\label{eq:propagation}
  \frac{\partial \psi}{\partial t} =&
  \nabla (D_{xx}\nabla \psi -\mathbf{V}_{c} \psi)
  +\frac{\partial}{\partial p}p^{2} D_{pp}\frac{\partial}{\partial p} \frac{1}{p^{2}}\psi
  -\frac{\partial}{\partial p} \left[ \dot{p} \psi -\frac{p}{3}(\nabla\cdot \mathbf{V}_{c})\psi \right]
  \nonumber \\
  & -\frac{1}{\tau_{f}}\psi
  -\frac{1}{\tau_{r}}\psi
  +q(\mathbf{r},p)  ,
\end{align}
where $\psi(\mathbf{r},p,t)$ is  the number density per unit of total particle momentum, which is related to the phase space density $f(\mathbf{r},p, t)$ as $\psi(\mathbf{r},p,t)=4\pi p^{2}f(\mathbf{r},p,t) $.
$D_{xx}$ is the spatial diffusion coefficient parametrized as
\begin{align}
\label{eq:Dxx}
D_{xx}=\beta D_{0} \left( \frac{\rho}{\rho_{0}} \right)^{\delta_{1,2}}  ,
\end{align}
where $\rho=p/(Ze)$ is the rigidity of the CR particles, and $\delta_{1(2)}$ is the index below (above) a reference rigidity $\rho_{0}$.  
The parameter $D_{0}$ is a normalization constant and $\beta=v/c$ is the ratio of the velocity $v$ of the CR particles to the speed $c$ of light.
$\mathbf{V}_{c}$ is the convection velocity related to the drift of CR particles from the Galactic disc due to the Galactic wind. 
The  diffusion in momentum space is described by  the re-acceleration parameter $D_{pp}$ related to the  Alfv$\grave{\mbox{e}}$n speed $V_{a}$ i.e. the velocity of turbulences in the hydrodynamical plasma, whose level is characterized as $\omega$\cite{Ginzburg:1990sk,Seo:1994aug}:
\begin{align}
D_{pp}=
\frac{4V_{a}^{2} p^{2}}
{3D_{xx}\delta_{i}
\left(4-\delta_{i}^{2}\right)
\left(4-\delta_{i}\right)\omega},
\end{align}
where $\delta_{i}=\delta_{1}$ or $\delta_{2}$ is the index of the spatial diffusion coefficient. 
$\dot{p}$, $\tau_{f}$ and $\tau_{r}$ are the momentum loss rate, the time scales for fragmentation and the time scales for radioactive decay, respectively.
The momentum loss rate of CR electrons is not the same as CR nucleons, and the relevant expressions are found in the APPENDIX C of paper \cite{Strong:1998pw}.

The source $q(\mathbf{r},p) $ of the primary particles is described as a broken power law spectrum multiplied by the assumed spatial distribution\cite{Strong:1998pw}:
\begin{align}
\label{eq:source}
q_{A}(R,z)
=
q_{0}c_{A}
\left( 
\frac{\rho}{\rho_{br}}
\right)^{\gamma_{s}}
 \left( \frac{R}{R_{\odot}} \right)^{\eta}
\exp
\left[
-\xi \frac{R-R_{\odot}}{R_{\odot}}
-\frac{|z|}{0.2~\text{kpc}}
 \right]~,
\end{align}
where $\eta=0.5$, $\xi=1.0$ and the parameter $q_{0}$ is normalized with the propagated flux. $c_{A}$ is the relative abundance of the Ath nucleon.
The reference rigidity $\rho_{br}$ is described as the breaks of injection spectrum. $\gamma_{s}$ is the power indices below(above) a reference rigidity. 
The flux of secondary particles is derived from the primary particle' s spectrum, spatial distribution and interaction with the interstellar gas.The calculation of secondary particle flux is referred the papers\cite{Strong:1998pw,Kelner:2006tc}. 

In this paper, the CR propagation equation \eqref{eq:propagation} is solved by GALPROP v54 package, which is based on a Crank-Nicholson implicit second-order scheme\cite{Strong:1998pw}.
In order to solve the equation, a cylindrically symmetric geometry is assumed. And the spatial boundary conditions assume that the density of CR particles vanishes at the boundaries of radius $R_h$ and half-height $Z_h$. 

At the top of the atmosphere of the Earth, CR particles are affected  by solar winds and the heliospheric magnetic field. The force-field approximation is used to describe this effect and the solar modulation potential $\phi$ denotes the force field intensity\cite{Gleeson:1968zza}. In this paper, we take $\phi=0.70$ GV (best fit value based on Model A in next section) and ignore the difference of $\phi$ between the experimental data.

In the three components of Galactic diffuse gamma ray emission, the $\pi^0$-decay component is calculated with the simulation of inelastic $p-p$ collisions producing secondary particles\cite{Strong:1998pw,Kelner:2006tc}, whose spectrum is mainly derived from the propagated CR protons. 
The IC component is calculated by the appropriate formalism based on the spatial and angular distribution of ISRF in the GALPROP code\cite{Strong:1998fr}. 
The Bremsstrahlung component is mainly from the contribution of CR electrons and positrons, whose calculation is referred the paper\cite{Strong:1998fr}. 

In this paper, the calculations of the CR propagation and the diffuse gamma ray emission are cross-checked with the results from the GALPROP webrun \cite{Vladimirov:2010aq}.
\section{Data selection and fitting schemes}\label{sec:fitSchemes}
In order to analyze the different constraints by the experimental data, the combinations of AMS-02, HESS and Milagro data are divided into the four types, which are denoted by model A, B, C and D, respectively. 
Model A, B and C are used to analyze CR elections constrained from the data of AMS-02, HESS and Milagro.
Model D is set up to analyze the $\pi^0$-decay component of diffuse gamma rays with AMS-02 protons, in which, source item of primary electrons is irrelevant and referred as model B.

The numerical solution of CR propagation equation \eqref{eq:propagation} is performed by GALPROP\cite{Strong:1998fr} package in the conventional re-accelebration diffusion model, which does not involve the  convection of CR particles on the Galactic disk.
In our previous paper, it is found that the propagation parameters: half-height $Z_h$,  diffuse parameters $D_{0}$ and $\delta(\delta=\delta_{1,2})$, and Alfv$\grave{\mbox{e}}$n speed $V_{a}$, and power indices:  $\gamma_{1,2}^{p}$ below(above) a reference rigidity of CR protons can be determined alone by the AMS-02 data: proton flux (P) and the ratio of Boron to Carbon flux (B/C)\cite{Jin:2014ica}. 
In this paper, besides these parameters, the normalization constant $N_p$ of CR protons and the parameters concerned with primary electrons are added too.

The source item \eqref{eq:source} of primary electrons in GALRPOP is described by use of the normalization constant $N_e$, the two reference rigidities(breaks) $\rho^{e}_{br1,2}$ and the three power indices $\gamma_{1,2,3}^{e}$.
The defaults of first and second reference rigidities are at 4 GV and 1 PV respectively in the conventional model of GALPROP. 
In this paper, as the maximal energy of GALRPOP grids is fixed at 100 TeV, the 1 PeV break of CR electron spectrum makes the third index invalid.
Thus, the models with 1 PV subscript describe CR electron spectrum of a simple power law and have no breaks above GeV  in the calculations of GALRPOP.    
In this paper, as the spectrum of CR electrons is focused above GeV, the first  reference rigidity $\rho^{e}_{br1}$ is not used in the fitting schemes. 
Based on the second reference rigidity and the two power indices below(above) the reference rigidity, we divide the fitting schemes into the three cases:  two free breaks, one free break and the fixed break. 
Two free breaks mean the two reference rigidities and the three power indices are free. One free break means the default of second reference rigidity is at 1 PV. The fixed break means the second reference rigidity is 2 TV from the feature of HESS electron spectrum.  

The CR electrons of AMS-02 involves the primary and secondary components theoretically. The primary component is not measured directly by experimental instruments. In order to analyze the relevant experimental value, the primary component of AMS-02 electron data may be extracted from the difference between AMS-02 electrons and positrons based on the hypothetic models.
In our previous paper, we gave the conclusion that the interpretation of  positron excess favors the annihilation of dark matter or the charge symmetry decay of dark matter\cite{Jin:2013nta}.
In this paper, generally, we assume that the sources relevant to the positron excess produce the charge symmetry particles in the final state, which means the flux of CR positrons is the same as CR electrons in excess of the background. Thus, the difference between the experimental data of CR electrons and positrons represents the measurement value of primary electrons. 
Based on this assumption, the fitting schemes are divided into two groups. The first group is denoted by $e^{-}$-$e^{+}$ case on the difference of AMS-02 electrons and positrons, and the second group is denoted by $e^{-}$ case on the AMS-02 electrons. 

 \begin{table}[htb]
\begin{center}
\begin{tabular}{l|rrr|rr|rr}
  \hline\hline
Exp.(\mbox{N}${}_{i}$)&\multicolumn{3}{l|}{AMS-02(72, 67, 72)}&\multicolumn{2}{l|}{HESS(8, 10)}\\
&$\chi^{2}_{\scriptsize\mbox{P}}$&$\chi^{2}_{\scriptsize\mbox{B/C}}$&$\chi^{2}_{\scriptsize\mbox{$e^{-}$ - $e^{+}$}}$&$\chi^{2}_{\scriptsize\mbox{$e^{\pm}_{08}$}}$&$\chi^{2}_{\scriptsize\mbox{$e^{\pm}_{09}$}}$
&\bf{$\chi^{2}_{\scriptsize\mbox{total}}$}&\bf{$\chi^{2}$/\scriptsize\mbox{N}}\\
\hline
A&90.80&76.29&38.93&&&206.02&0.98\\
A${}_{\scriptsize\mbox{2TV}}$&97.80&69.98&72.64&&&240.42&1.14\\
A${}_{\scriptsize\mbox{1PV}}$&97.70&70.06&71.05&&&238.81&1.13\\
B&90.86&76.23&40.41&13.93&19.72&241.16&1.05\\
B${}_{\scriptsize\mbox{2TV}}$&102.32&66.84&72.95&69.29&62.83&374.22&1.63\\
B${}_{\scriptsize\mbox{1PV}}$&102.54&66.70&71.54&35.98&55.95&332.71&1.45\\
\hline\hline
\end{tabular}
\end{center}
\caption{The best-fit  $\chi^2$ relevant to the $e^{-}$-$e^{+}$ case in model A-B. 1 PV subscript means the spectrum of primary electrons is a simple power law. 2 TV subscript means the spectrum of CR electrons have a 2TV break\cite{Ackermann:2010ij}. the N${}_{i}$ is number of the experiment data points and as an example, the number of AMS-02 proton data points is 72. The total $\chi^2$ and its value over the total data-points  of the chosen experiment of each model are listed in the two tail columns.}
\label{tab:chisquares_electronMinusPositon}
\end{table}

\begin{table}[htb]
\begin{center}
\begin{tabular}{l|rrr|rr|rr|rr}
  \hline\hline
Exp.(\mbox{N}${}_{i}$)&\multicolumn{3}{l|}{AMS-02(72, 67, 73)}&\multicolumn{2}{l|}{HESS(8, 10)}&\multicolumn{2}{l|}{Milagro(1, 1)}&&\\
&$\chi^{2}_{\scriptsize\mbox{P}}$&$\chi^{2}_{\scriptsize\mbox{B/C}}$&$\chi^{2}_{\scriptsize\mbox{$e^{-}$}}$&$\chi^{2}_{\scriptsize\mbox{$e^{\pm}_{08}$}}$
&$\chi^{2}_{\scriptsize\mbox{$\gamma^{s1}$}}$&$\chi^{2}_{\scriptsize\mbox{$\gamma^{s2}$}}$&\bf{$\chi^{2}_{\scriptsize\mbox{total}}$}&\bf{$\chi^{2}$/\scriptsize\mbox{N}}\\
\hline
A&91.04 &75.98 &42.77 && &&&209.79 &0.99\\
A${}_{\scriptsize\mbox{2TV}}$&107.64 &63.60 &130.78 && &&&302.03 &1.43\\
A${}_{\scriptsize\mbox{1PV}}$&107.45 &63.71 &126.97 && &&&298.13 &1.41\\
B&91.02 &76.02 &42.91 &13.58 &13.65 &&&237.17 &1.04\\
B${}_{\scriptsize\mbox{2TV}}$&113.71 &60.59 &129.49 &56.75 &50.01 &&&410.55 &1.79\\
B${}_{\scriptsize\mbox{1PV}}$&112.82 &61.01 &125.77 &24.71 &42.60 &&&366.90 &1.60\\
C&90.66 &76.86 &42.79 &13.81 &13.47 &5.04 &5.69 &248.32 &1.07\\
D&91.28 &76.10 &&&&5.10 &5.76 &178.24 &1.26\\
 \hline\hline
\end{tabular}
\end{center}
\caption{The same as Table \ref{tab:chisquares_electronMinusPositon}, but for the $e^{-}$ case in model A-D.
$\gamma^{s1}$ and $\gamma^{s2}$ denote the Galactic sky $l\in[30^\circ,65^\circ]$ and $l\in[65^\circ,85^\circ]$.
The added model C and D are used to limit the higher energy spectrum of CR electrons and protons than the HESS data.
}
\label{tab:chisquares_electron}
\end{table}
 
In summary, in the fitting schemes, besides the parameters: half-height $Z_h$, etc. of context, the normalization constant $N_e$, the second reference rigidity $\rho^{e}_{br2}$ and the three power indices $\gamma_{1,2,3}^{e}$ of primary electron are included. In total, there are 12 fitting parameters.
In the chosen groups of experimental data, AMS-02 protons, AMS-02 B/C,  AMS-02 electrons(or the difference of AMS-02 electrons and positrons), HESS electrons released in 2008, HESS electrons in 2009, Milagro $l\in[30^\circ,65^\circ]$ diffuse gamma ray and  Milagro $l\in[65^\circ,85^\circ]$ diffuse gamma ray are denoted by P, B/C, $e^{-}$(or $e^{-}$-$e^{+}$), $e^{\pm}_{08}$, $e^{\pm}_{09}$, $\gamma^{s1}$ and $\gamma^{s2}$, respectively. In total, there are 7 groups of data.
In model A and B,  the second reference rigidity of 2 TV and 1 PV are denoted by model A(B)${}_{\scriptsize\mbox{2TV}}$ and  A(B)${}_{\scriptsize\mbox{1PV}}$. Because of the only considering the TeV scale, 1PV case means the only one reference rigidity $\rho^{e}_{br1}$ available in the CR electron source item \eqref{eq:source} and simple power law spectrum of the primary electron.
The parameters and models are shown in Table  \ref{tab:para_electronMinusPositon}-\ref{tab:para_electron}.

Through the global $\chi^2$-fit using the MINUIT package, the best-fit values of the parameters and spectrum of CRs are derived from the minimized $\chi^2$. As an example, the corresponding relations of the models A-D and their experimental data concerned are found in Table \ref{tab:chisquares_electron}, which shows the best-fit  $\chi^2$ of the models A-D and relevant experiments. The 12 parameters of each model in the models A-D are shown in Table \ref{tab:para_electron}, which shows the best-fit parameters of each model.  
\begin{figure}
\includegraphics[width=0.49\textwidth]{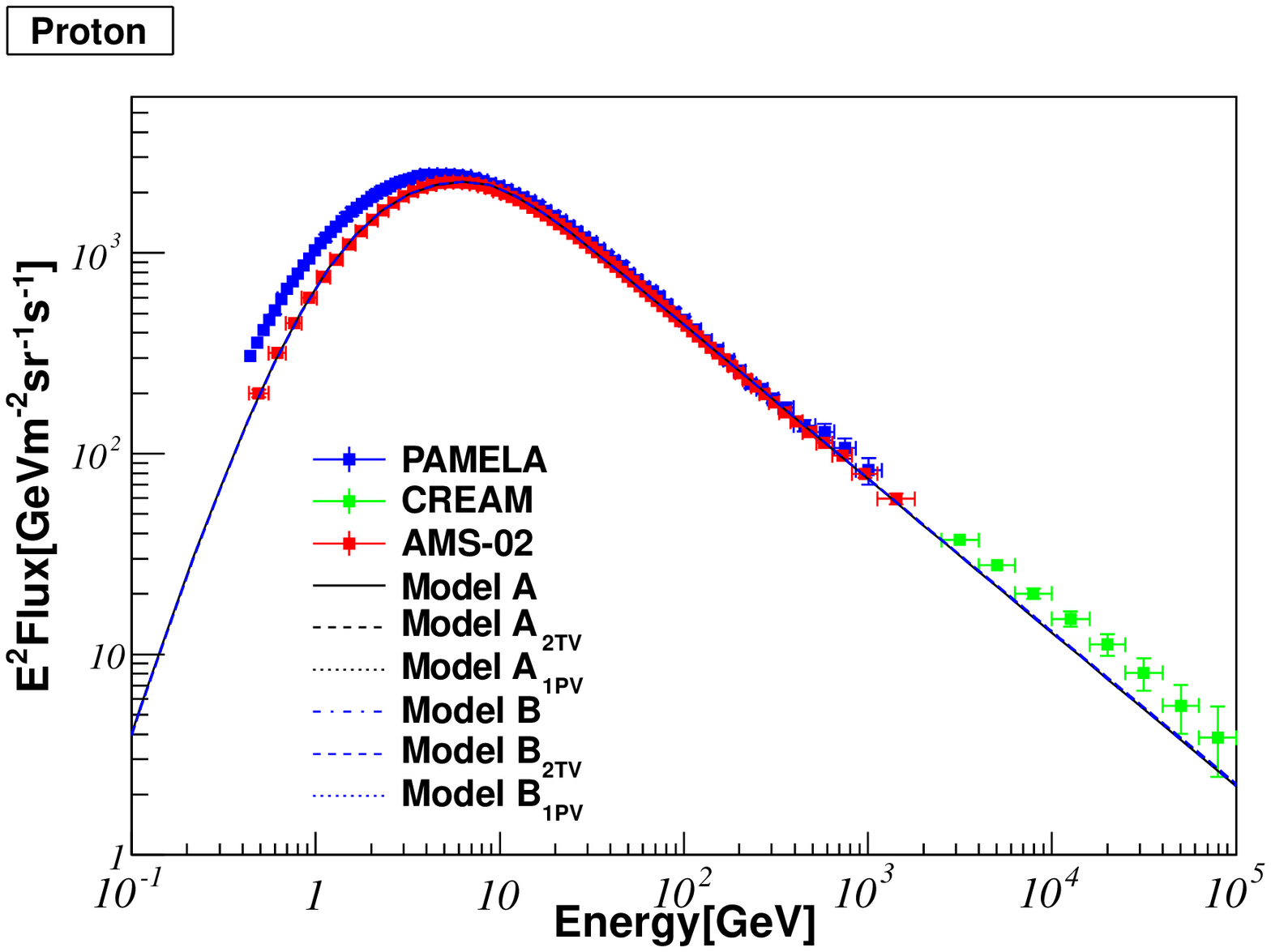}\includegraphics[width=0.49\textwidth]{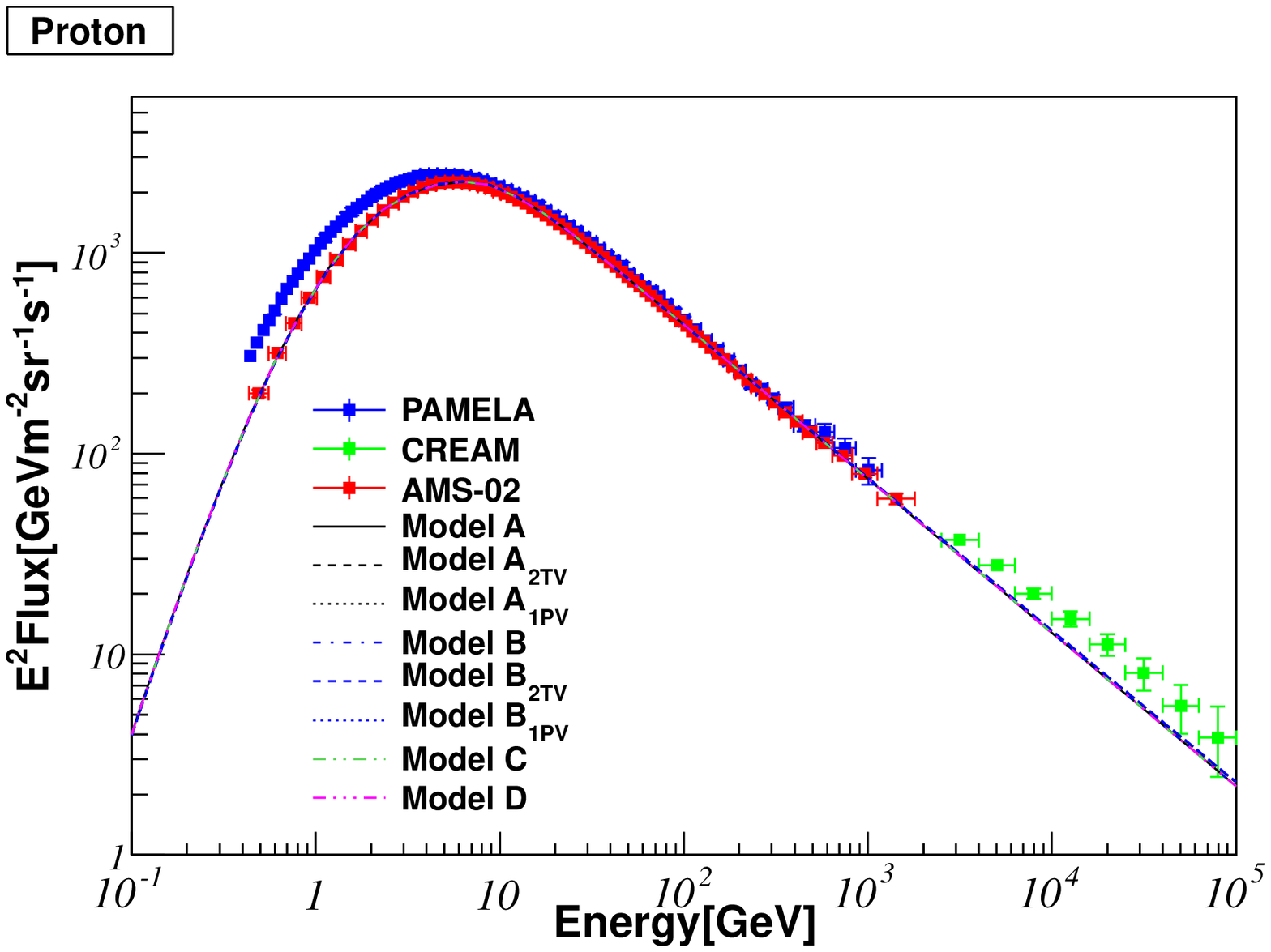}
\includegraphics[width=0.49\textwidth]{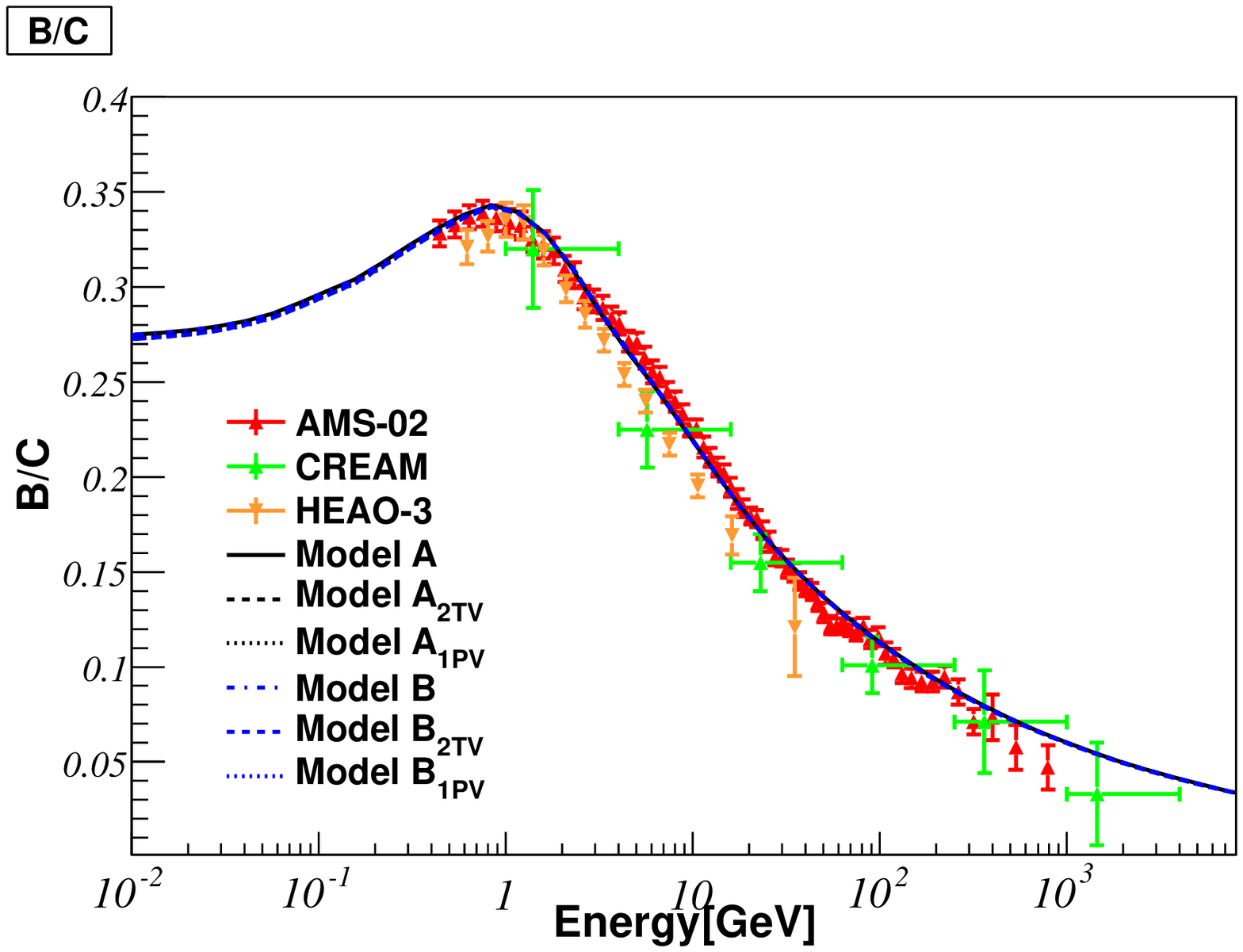}\includegraphics[width=0.49\textwidth]{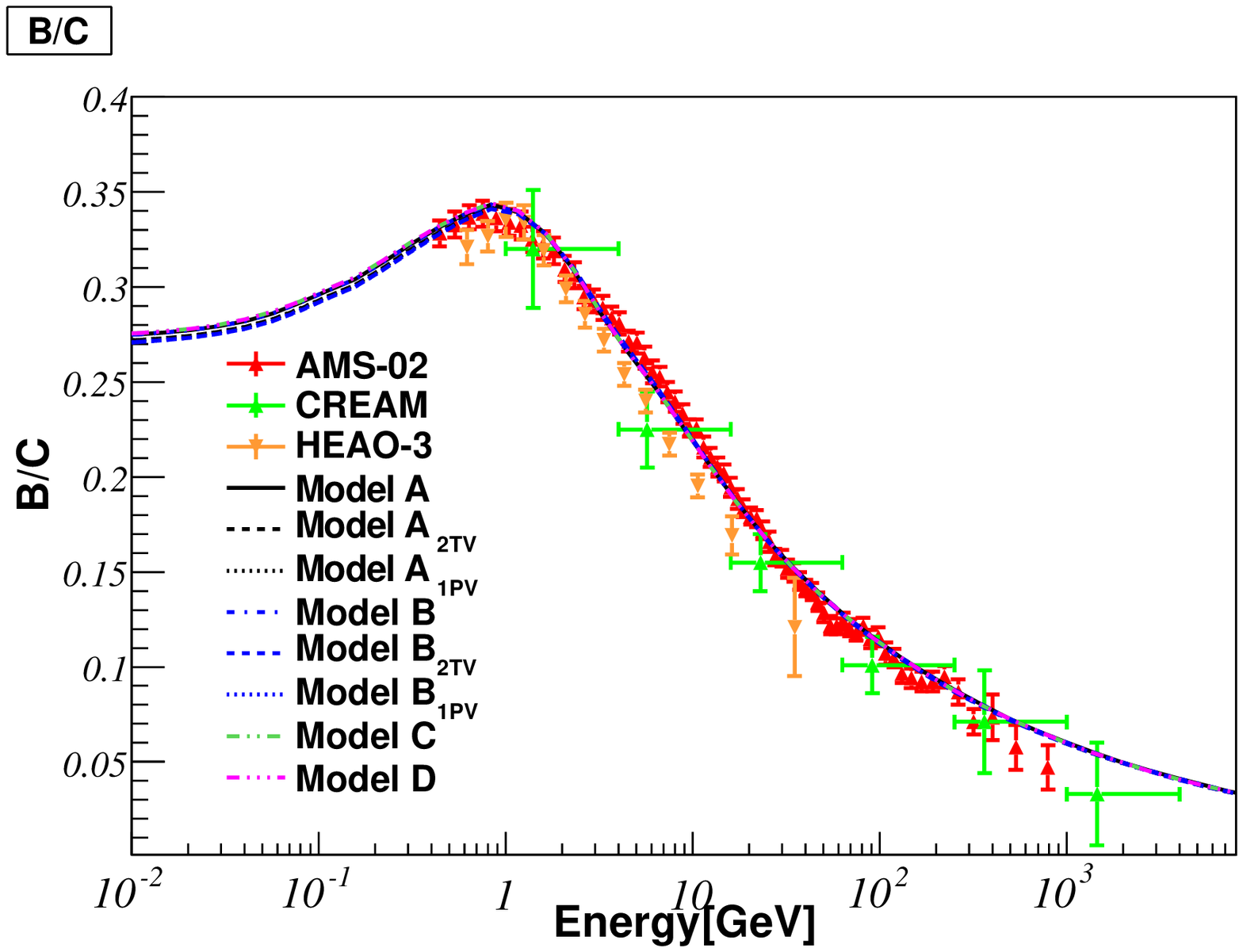}
\includegraphics[width=0.49\textwidth]{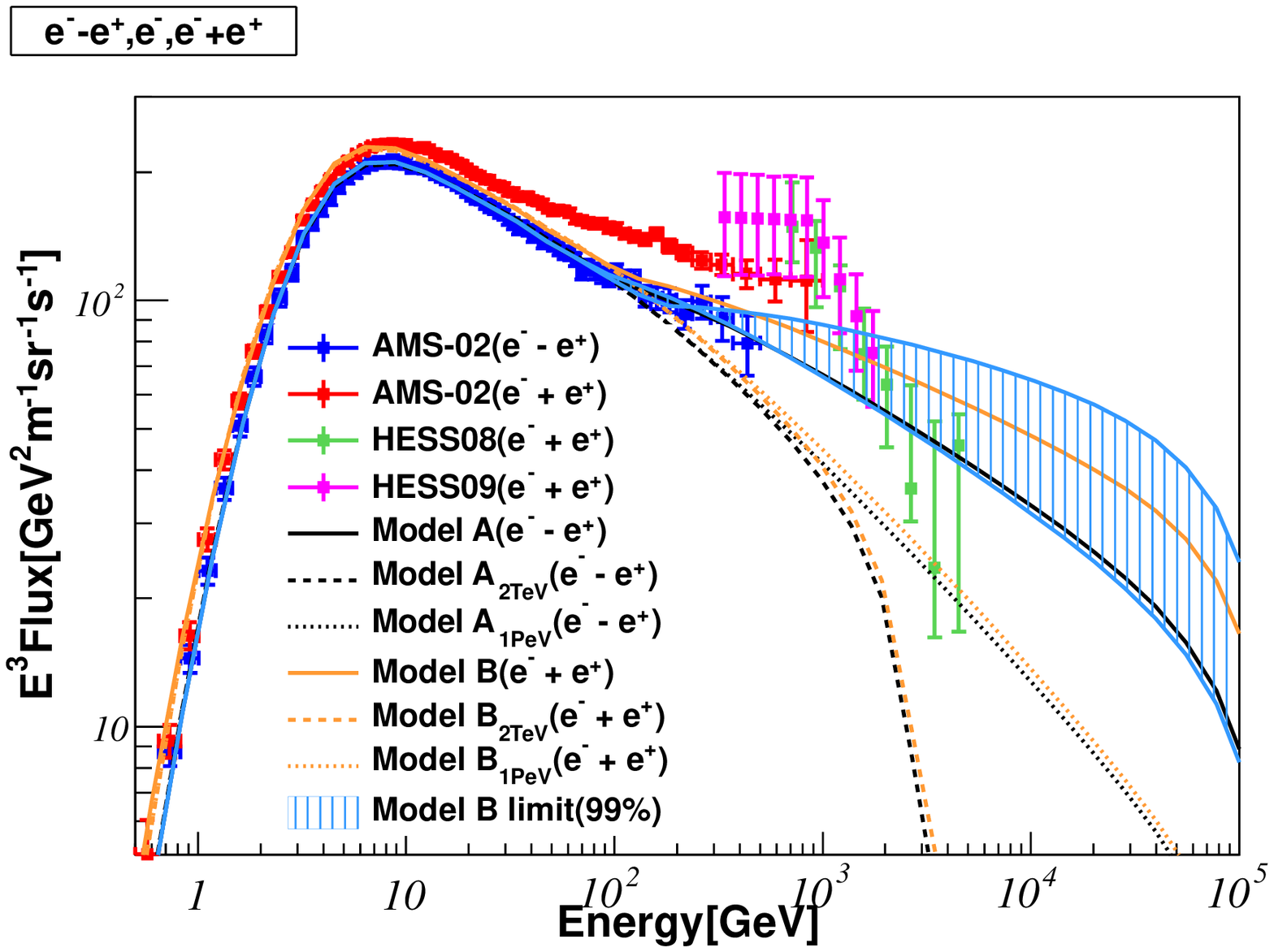}\includegraphics[width=0.49\textwidth]{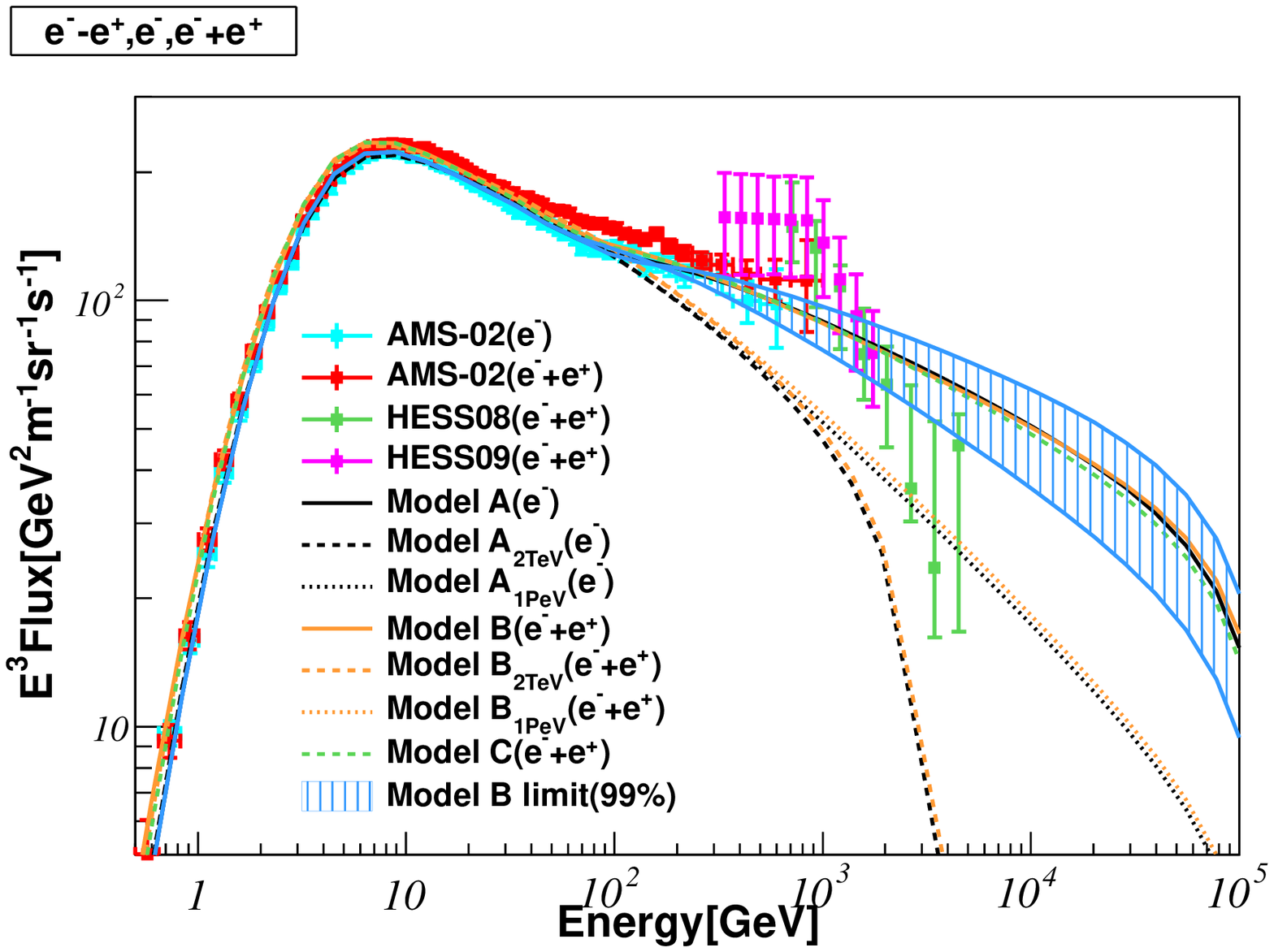}
\caption
{(Left column) $e^{-}$-$e^{+}$ case in model A-B. (Right column) $e^{-}$ case in model A-D. 
(First row) the flux of CR Protons relevant to the models and of the measurements from AMS-02\cite{Aguilar:2015ooa}, PAMELA\cite{Adriani:2011cu} and and CREAM\cite{Yoon:2011aa} experiments.
(Second row)  the ratio of Boron to Carbon flux relevant to the models and of the measurements from  AMS-02\cite{oliva:2015BC}, HEAO-03\cite{Engelmann:1990zz}, and CREAM\cite{Ahn:2008my} experiments.
(Last row)the flux of CR electrons relevant to the models and of the measurements from AMS-02\cite{Aguilar:2014mma,Aguilar:2014fea} and HESS\cite{Aharonian:2008aa,Aharonian:2009ah} experiments.  $e^{-}$-$e^{+}$ denoted by the difference between the flux of CR electrons and positrons from AMS-02\cite{Aguilar:2014mma}. In the legends of last row figures, $e^{-}$+$e^{+}$ means the lines is relevant to the total fluxes of CR electrons and positrons. 
The filling regions are drawn from the maximal and minimal fluxes of primary electrons and CR electrons calculated on the parameters of Table \ref{tab:para_MinMax}. 
}
 \label{fig:AMS02_ProtonAndBC}
\end{figure}

\begin{table}[htb]
\begin{center}
\begin{tabular}{lrrrrrrrrr}
\hline\hline
Para.&A&A${}_{\scriptsize\mbox{2TV}}$&A${}_{\scriptsize\mbox{1PV}}$&B&B${}_{\scriptsize\mbox{2TV}}$&B${}_{\scriptsize\mbox{1PV}}$\\
\hline
$N_e$&1.287 &1.299 &1.299 &1.287 &1.300 &1.299\\
$\gamma^{e}_{1}$&1.607 &1.632 &1.632 &1.607 &1.640 &1.640\\
$\gamma^{e}_{2}$&2.750 &2.723 &2.723 &2.750 &2.715 &2.715\\
$\rho^{e}_{br2}$&0.108 &2 &1000 &0.128 &2 &1000\\
$\gamma^{e}_{3}$&2.520 &5 &5 &2.440 &5 &5\\
$\gamma^{e}_{3}$- $\gamma^{e}_{2}$&\textcolor{red}{\bf{-0.230}}&2.277 &&\textcolor{red}{\bf{-0.310}}&2.285 &&\\
\hline
$Z_h$&2.953 &2.960 &2.959 &2.953 &2.954 &2.957\\
$D_0$&1.751 &1.757 &1.757 &1.752 &1.762 &1.762\\
$\delta$&0.296 &0.296 &0.296 &0.296 &0.296 &0.296\\
$v_{Alfven}$&38.61 &39.07 &39.07 &38.62 &39.37 &39.39\\
\hline
$N_p$&4.523 &4.520 &4.520 &4.523 &4.518 &4.518\\
$\gamma^{p}_{1}$&1.779 &1.775 &1.775 &1.779 &1.774 &1.774\\
$\gamma^{p}_{2}$&2.468 &2.464 &2.464 &2.468 &2.463 &2.463\\
\hline\hline
\end{tabular}
\end{center}
\caption{Best-fit values of the parameters relevant to the $e^{-}$-$e^{+}$ case in model A-B. 1 PV subscript means the spectrum of primary electrons is a simple power law. 2 TV subscript means the spectrum of CR electrons have a 2TV break\cite{Ackermann:2010ij}. The units of $N_e$ and $N_p$ are both $10^{-9}\mbox{MeV}^{-1}\mbox{cm}^{-2}\mbox{sr}^{-1}\mbox{s}^{-1}$.  $\rho^{e}_{br2}$, $Z_h$, $D_0$ and $v_{Alfven}$ are in units of TV, kpc, $10^{28}\mbox{cm}^2\mbox{s}^{-1}$ and km $\mbox{s}^{-1}$. dZ, vertical height Z's bin in the grid of Galactic disk, is modified as 0.2kpc to reduce the computing times. The rest parameters of GALPROP is referred to the example 01 of WebRun\cite{Vladimirov:2010aq}. In the numbers, the red ones are negative, which mean the spectrum of CR electrons begins to harden across the reference break. }
\label{tab:para_electronMinusPositon}
\end{table}

\begin{table}[htb]
\begin{center}
\begin{tabular}{lrrrrrrrrr}
  \hline\hline
Para.&A&A${}_{\scriptsize\mbox{2TV}}$&A${}_{\scriptsize\mbox{1PV}}$&B&B${}_{\scriptsize\mbox{2TV}}$&B${}_{\scriptsize\mbox{1PV}}$&C&D\\
\hline
$N_e$&1.360 &1.378 &1.378 &1.360 &1.378 &1.378 &1.360 &1.360\\
$\gamma^{e}_{1}$&1.613 &1.654 &1.653 &1.613 &1.660 &1.659 &1.614 &1.613\\
$\gamma^{e}_{2}$&2.730 &2.683 &2.683 &2.730 &2.675 &2.677 &2.731 &2.730\\
$\rho^{e}_{br2}$&0.095 &2 &1000 &0.091 &2 &1000 &0.091 &0.091\\
$\gamma^{e}_{3}$&2.457 &5 &5 &2.470 &5 &5 &2.468 &2.470\\
$\gamma^{e}_{3}$- $\gamma^{e}_{2}$&\textcolor{red}{\bf{-0.273}} &2.317 &&\textcolor{red}{\bf{-0.260}} &2.325 &&\textcolor{red}{\bf{-0.262}} &\textcolor{red}{\bf{-0.260}}\\
\hline
$Z_h$&2.952 &2.954 &2.956 &2.953 &2.955 &2.955 &2.957 &2.955\\
$D_0$&1.750 &1.766 &1.766 &1.750 &1.773 &1.772 &1.751 &1.744\\
$\delta$&0.296 &0.296 &0.296 &0.296 &0.296 &0.296 &0.296 &0.297\\
$v_{Alfven}$&38.61 &39.65 &39.64 &38.61 &39.00 &39.95 &38.58 &38.48\\
\hline
$N_p$&4.523 &4.516 &4.516 &4.523 &4.514 &4.515 &4.527 &4.527\\
$\gamma^{p}_{1}$&1.779 &1.772 &1.772 &1.779 &1.771 &1.771 &1.781 &1.779\\
$\gamma^{p}_{2}$&2.467 &2.461 &2.461 &2.467 &2.459 &2.460 &2.468 &2.467\\
\hline\hline
\end{tabular}
\end{center}
\caption{The same as Table \ref{tab:para_electronMinusPositon}, but for the $e^{-}$ case in model A-D. The added model C and D are used to limit the higher energy spectrum of CR electrons and protons than the HESS data.}
\label{tab:para_electron}
\end{table}

\section{Results}\label{sec:results}
\begin{figure}
\includegraphics[width=0.49\textwidth]{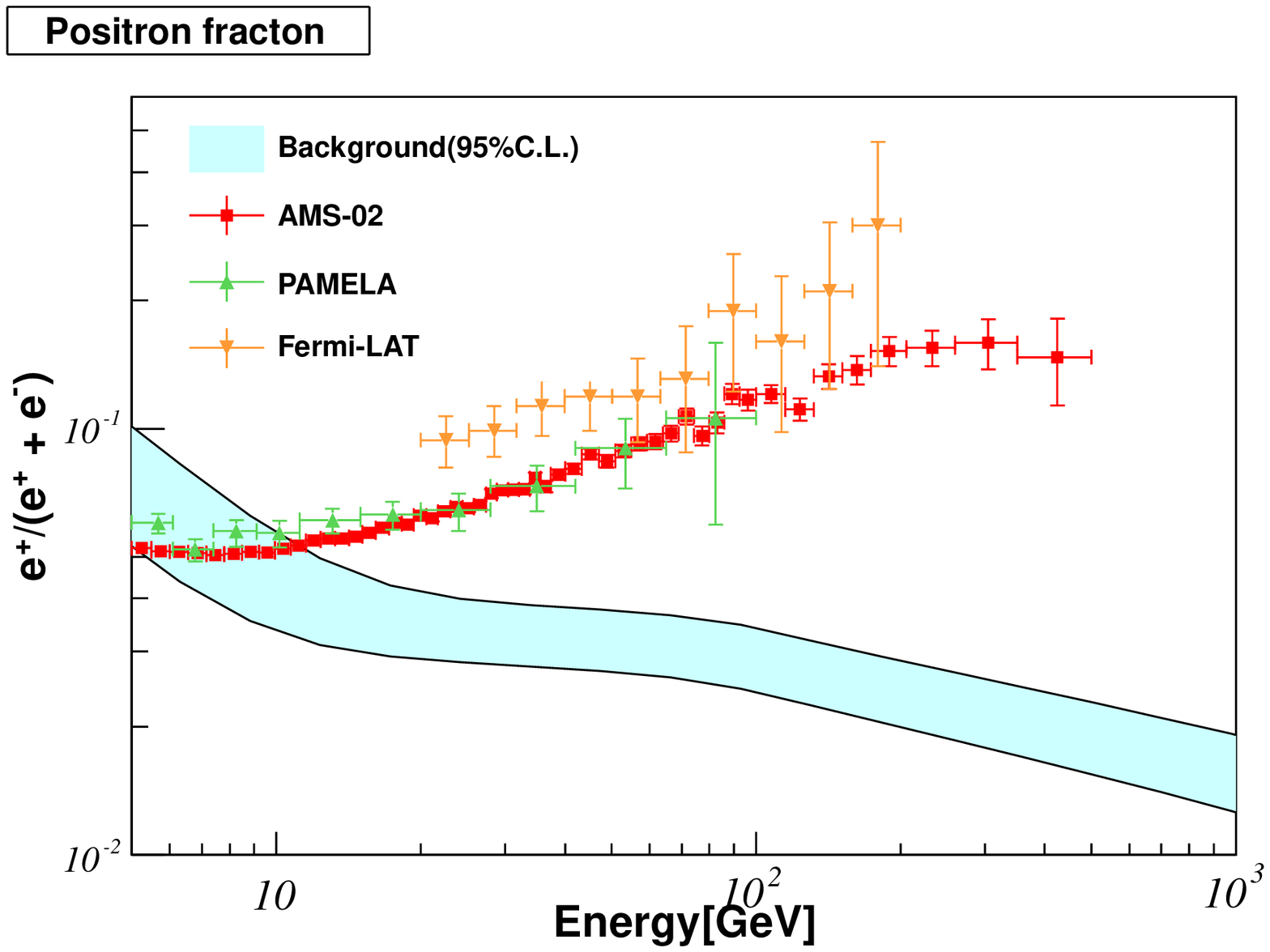}\includegraphics[width=0.49\textwidth]{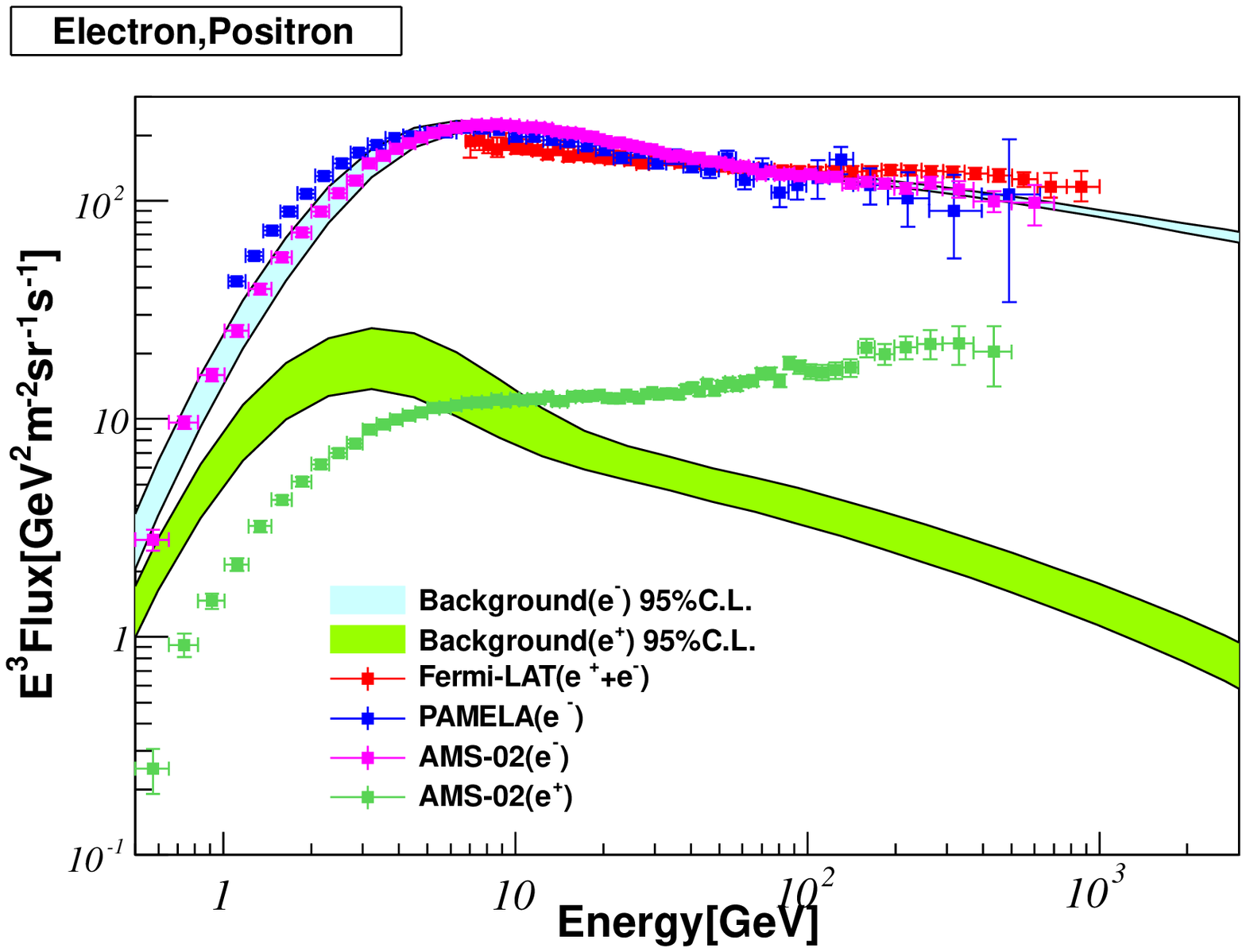}
\caption{ Predictions for positron fraction (left) and fluxes of CR electrons and positrons (right) in the background with the uncertainties of propagation parameters within $95\%$ C.L.. For positron fraction, the data of AMS-02\cite{Accardo:2014lma} PAMELA \cite{Adriani:2010ib} and Fermi-LAT \cite{FermiLAT:2011ab} are  shown.
For electron and positron fluxes, the data of PAMELA (electrons) \cite{Adriani:2011xv}, Fermi-LAT (electrons+positrons) \cite{Ackermann:2010ij} and AMS-02 (electrons and positrons)
\cite{Aguilar:2014mma} are also shown. The propagation parameters with errors are referred the paper \cite{Jin:2014ica}. }
 \label{fig:PBC_EPBG}
\end{figure}

In Table \ref{tab:chisquares_electronMinusPositon} and \ref{tab:chisquares_electron}, the best-fit $\chi^2$ relevant to  the $e^{-}$-$e^{+}$ case in model A-B and the $e^{-}$ case in model A-D are shown respectively. 
And the total data-points and the ratio of total $\chi^2$ to the total data-points in each model are listed in the last two columns.
In all of the models, best-fit $\chi^2$ relevant to the AMS-02 P and B/C is almost around the data-points of experimental data.
Thus, the theoretical CR proton flux and B/C are consistent with the AMS-02 data. 
As seen in Figure \ref{fig:AMS02_ProtonAndBC}, CR proton flux and B/C are not distinctly discriminated between models.
In Table \ref{tab:para_electronMinusPositon} and \ref{tab:para_electron}, the propagation parameters relevant to the best-fit $\chi^2$ have differences only within 5\% between the models.
It implies that the propagation parameters are strongly constrained by the high accuracy data of AMS-02 Proton and B/C and do not take the apparent uncertainties to the calculation of CR electrons and positrons. 
In our previous paper\cite{Jin:2014ica}, the uncertainties of propagation parameters are analyzed in great detail and relevant errors and contours at 95\% C.L. are shown in the  table 2 and figure 2 respectively. Based on the previous sampling data, the limited fluxes of CR electrons and positrons are predicted.  
In Figure \ref{fig:PBC_EPBG}, the positron fraction and fluxes of CR electrons and positrons with the uncertainties of propagation parameters at $95\%$ C.L.. are shown. 
From the limited regions, it is found that above 10GeV, the positron fraction and fluxes of CR positrons have errors around 30\% errors, and the errors of CR electron flux is less than 10\%.
For the $e^{-}$ case in model D, the best-fit  $\chi^2$ shows the flux of $\pi^0$-decay gamma rays, derived from CR protons, is not far less than the one from Milagro data in the errors of  AMS-02 data. 
As seen in the first rows of Figure \ref{fig:AMS02_ProtonAndBC}, the high energy flux of CREAM Protons matches the extension of  AMS-02 Protons, which means the CR protons in AMS-02 data are very difficultly fitting to the excess of diffuse gamma rays from Milagro experiments.  
\subsection{The spectra of CR electrons}
In the last rows of Figure \ref{fig:AMS02_ProtonAndBC}, the fluxes of CR electrons, the total of CR electrons and positrons, and the primary electrons are drawn. As seen in the Table \ref{tab:chisquares_electronMinusPositon} and \ref{tab:chisquares_electron},  in two cases of the model A-B,  the ratio of best-fit  $\chi^2$ to the data-points of AMS-02 electrons is all less than 2, which means the all models are consistent with AMS-02 electron data.

As $\chi^2$ relevant to the AMS-02 P and B/C has no significant difference between the models, and the fluxes of secondary positrons and electrons produced mainly from CR proton's interaction with ISM are almost equivalent to  each other, the spectral features of secondary positrons and electrons with a simple power law do not contribute to the changes of the primary electron structure in the constraint of these models. 

 \begin{figure}
\includegraphics[width=0.49\textwidth]{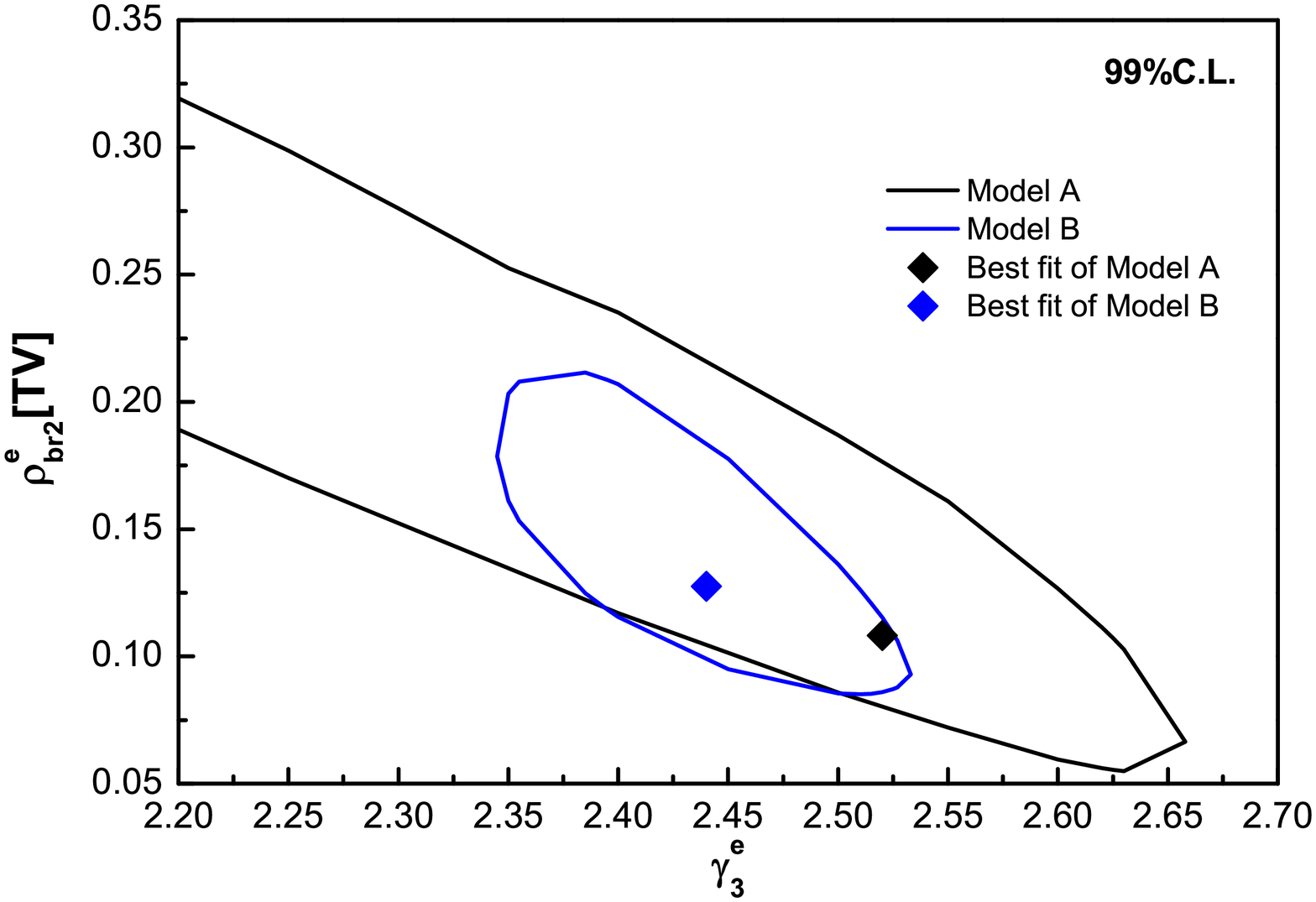}\includegraphics[width=0.49\textwidth]{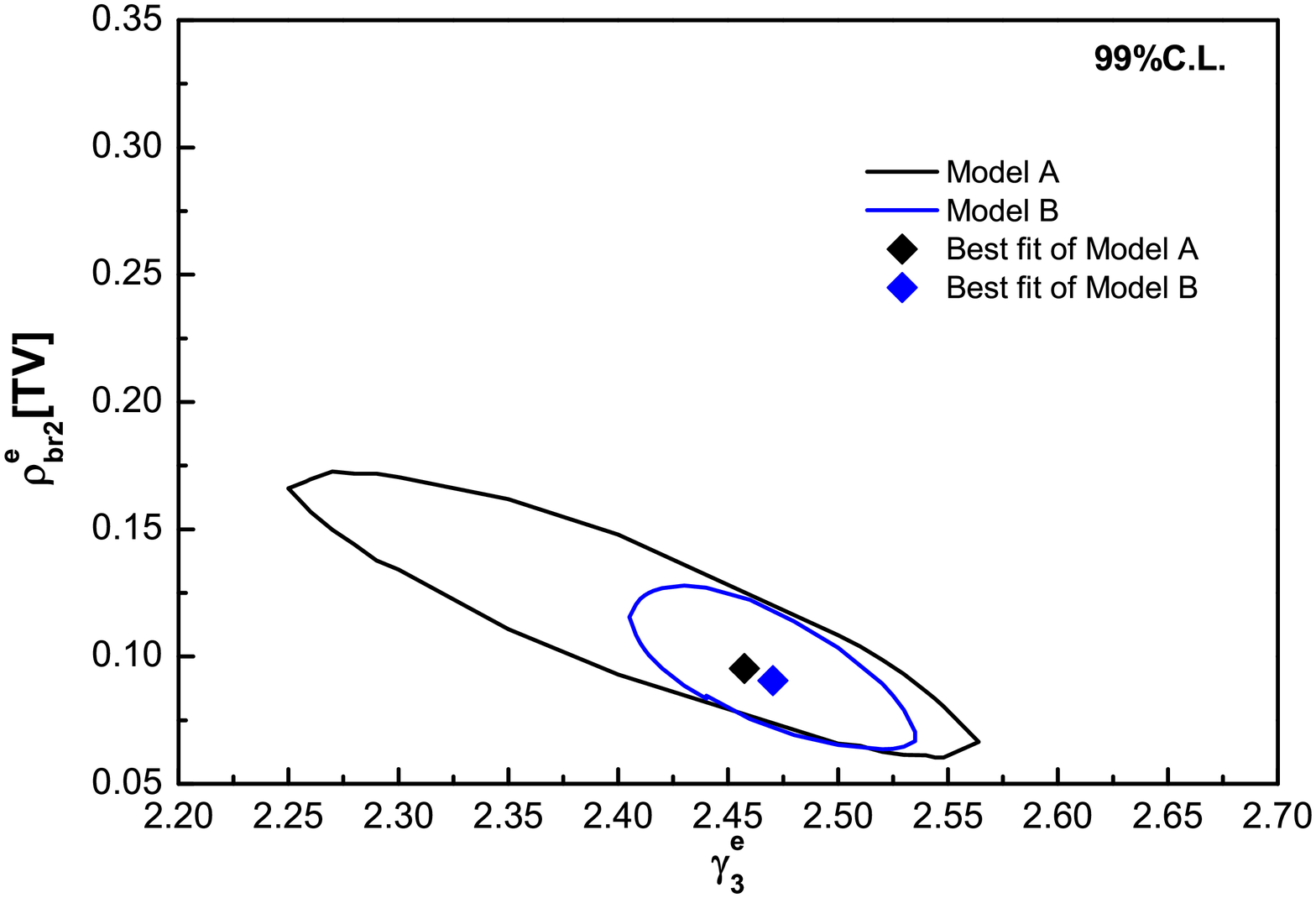}
\includegraphics[width=0.49\textwidth]{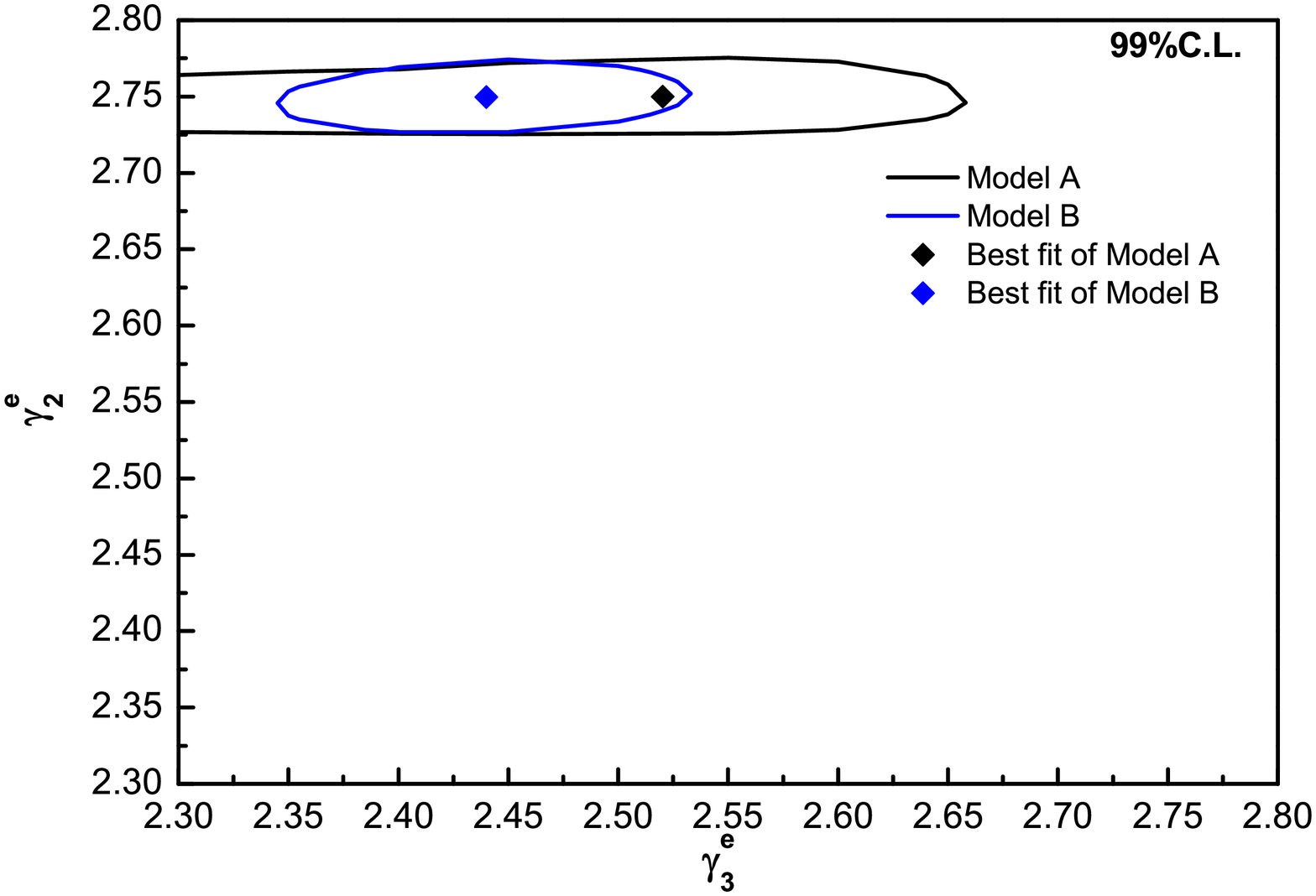}\includegraphics[width=0.49\textwidth]{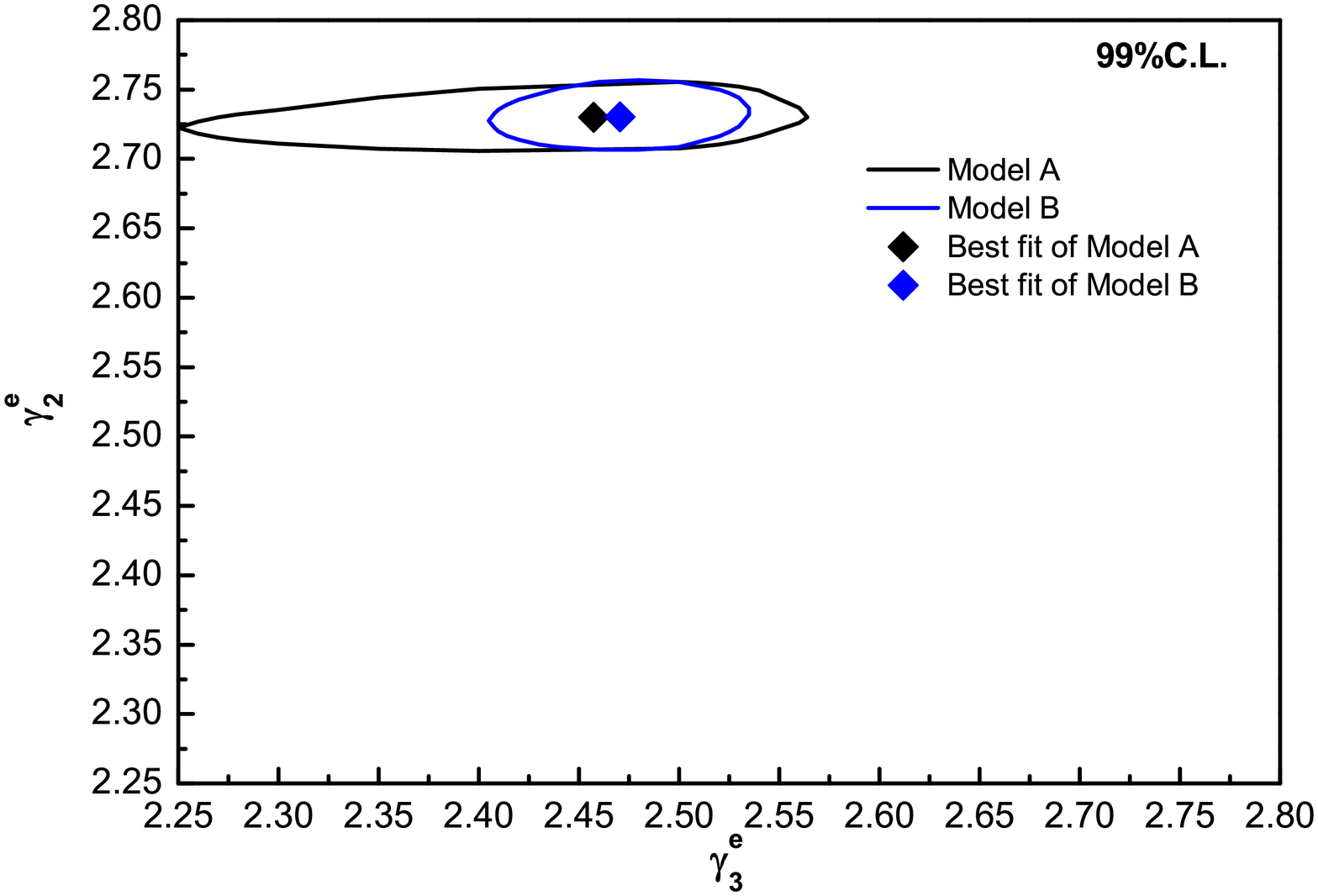}
\caption{
(First row) allowed regions in ($\gamma_{3}^{e}$, $\rho^{e}_{br2}$) plane at 99\% C.L. for (left) $e^{-}$-$e^{+}$ cases and (right)  $e^{-}$ cases in model A-B.
(Second row) allowed regions in ($\gamma_{3}^{e}$, $\gamma_{2}^{e}$) plane at 99\% C.L. for (left) $e^{-}$-$e^{+}$ cases and (right)  $e^{-}$ cases in model A-B.
}
 \label{fig:InjContour}
\end{figure}
In the $e^{-}$-$e^{+}$ and  $e^{-}$ cases of model A, the $\chi^2$ relevant to 1 PV break (means a simple power law spectrum) is around the data-points of  $e^{-}$-$e^{+}$ and  $e^{-}$ experimental data, which is seen in Table \ref{tab:para_electronMinusPositon} and \ref{tab:para_electron}. It is justified that the data of AMS-02 electrons implies the basic feature of the power law and the primary electrons are the dominant component of CR electrons. The maximal value of positron fraction from AMS-02 is less than 20\% and in line with this feature too. 
 
However, in details, the experimental data of CR electrons implies the possible structures recently, such as Fermi-LAT\cite{Ackermann:2010ij}, PAMELA\cite{Adriani:2011xv} and AMS-02\cite{Aguilar:2014mma}. 
Specially, the high accuracy data of  the latest AMS-02 electrons shows the fine features of spectra, which are described as the positron excess against the astrophysical background. 
In the $e^{-}$ cases, the $\chi^2$ relevant to model A are much less than the 1 PV break case of model A and justify these features of spectra. 
As seen in Table \ref{tab:para_electron}, most of the absolute differences of power indices above(below) the around 100GV reference rigidity are greater than 0.2 and shows the breaks of CR electron spectrum clearly. 
In the $e^{-}$-$e^{+}$ cases, it also implies that the structure of the primary electrons is not described by a simple power law. Thus, the positron excess against a power law spectrum do not justify the existence of the same flux of CR electron excess.
 
In order to explore the primary electron excess further, as the primary electrons and CR electrons of AMS-02  both do not exclude the 1 PV and 2TV break cases, it is necessary to include HESS data.
In the $e^{-}$-$e^{+}$ and  $e^{-}$ cases of model B, $\chi^2$ relevant to HESS electrons of model B are neither greater than the double data-points of HESS electrons and show the flux of HESS electrons match the TeV scale extension of AMS-02 electron flux. 
Nevertheless, the $\chi^2$ relevant to 2 TV and 1 PV breaks are both greater than the double data-points of HESS electrons. It is justified that and the primary electrons and CR electrons constrained by AMS-02 and HESS both do not favor the 2 TeV breaks or a simple power law. 

In order to explore the CR electron spectrum structure's limit to a power law, we do some analysis of confidence intervals to illustrate the changes of the second reference rigidity $\rho^{e}_{br2}$ and the two indices $\gamma_{2}^{e}$ and $\gamma_{3}^{e}$ in the source item of primary electrons.
In the first row of Figure \ref{fig:InjContour}, the allowed regions of the second reference rigidity and the third index in the $e^{-}$-$e^{+}$ cases and the $e^{-}$ cases in model A-B are shown at the 99\% C.L..
In the secondary row of Figure \ref{fig:InjContour}, the allowed regions of the second and third indices in the $e^{-}$-$e^{+}$ cases and the $e^{-}$ cases in model A-B are also shown at the 99\% C.L..

In the $e^{-}$-$e^{+}$ case of model A, the confidence intervals of third index are not constrained in the reasonable ranges but the maximum of third index only limited. 
From the left column of Figure \ref{fig:InjContour}, when the third index is maximum 2.674 at 99\%C.L., the second reference rigidity drops to 50GV and the second index is near 2.749.
The difference of 0.075 between the second and third indices at the reference 64GV means that from the constraint of AMS-02 data alone, the primary electron spectrum is not excluded from a simple power law at the 99\% C.L.. 
Nevertheless, in the $e^{-}$-$e^{+}$ case of model B, this situation is changed. The confidence intervals of third index and second reference rigidity are constrained in the narrower ranges.
At 99\% C.L. the third index decreases from 2.54 to 2.35, but the second ones are only in the range 2.746 - 2.751. Thus, the spectrum of primary electrons above the hundred GeV depends on whether or not to include HESS data and the spectrum of primary electrons favors the feature, which is more complex than a simple power law. Before the AMS-02 releases the measurement data of CR electrons, the paper\cite{Feng:2013zca} has predicted this feature theoretically. 
These results implies that spectrum of the primary electrons need the different interpretation from the conventional astrophysical background, such as the nearby SNRs\cite{DiMauro:2014iia}, asymmetric charge dark matter, etc. 
In the $e^{-}$ case of model A, the confidence intervals of the second and third indices and second reference rigidity are constrained completely by AMS-02 data alone.
As seen in the right column of  Figure \ref{fig:InjContour}, at the 99\% C.L., the difference between the second and third indices is significant. Thus, the theoretical CR electrons constrained by AMS-02 electrons have the obvious structure, which implies the components in excess of the astrophysical background.
For  $e^{-}$ case in model B, the difference between the second and third indices is much less than the model A. 
In the Table \ref{tab:para_MinMax}, the bounds of propagation parameters from the $e^{-}$-$e^{+}$ and $e^{-}$ cases of model B are listed. $B^{e^{-} - e^{+}}_{\scriptsize\mbox{Min}}$ and  $B^{e^{-} - e^{+}}_{\scriptsize\mbox{Max}}$ represent respectively the minimal and maximal flux of primary electrons in the $e^{-}$-$e^{+}$ case of model B.  $B^{e^{-}}_{\scriptsize\mbox{Min}}$ and  $B^{e^{-}}_{\scriptsize\mbox{Max}}$ are the same as  $e^{-}$-$e^{+}$ case, but for $e^{-}$ case of model B. 

\begin{figure}
\includegraphics[width=0.49\textwidth]{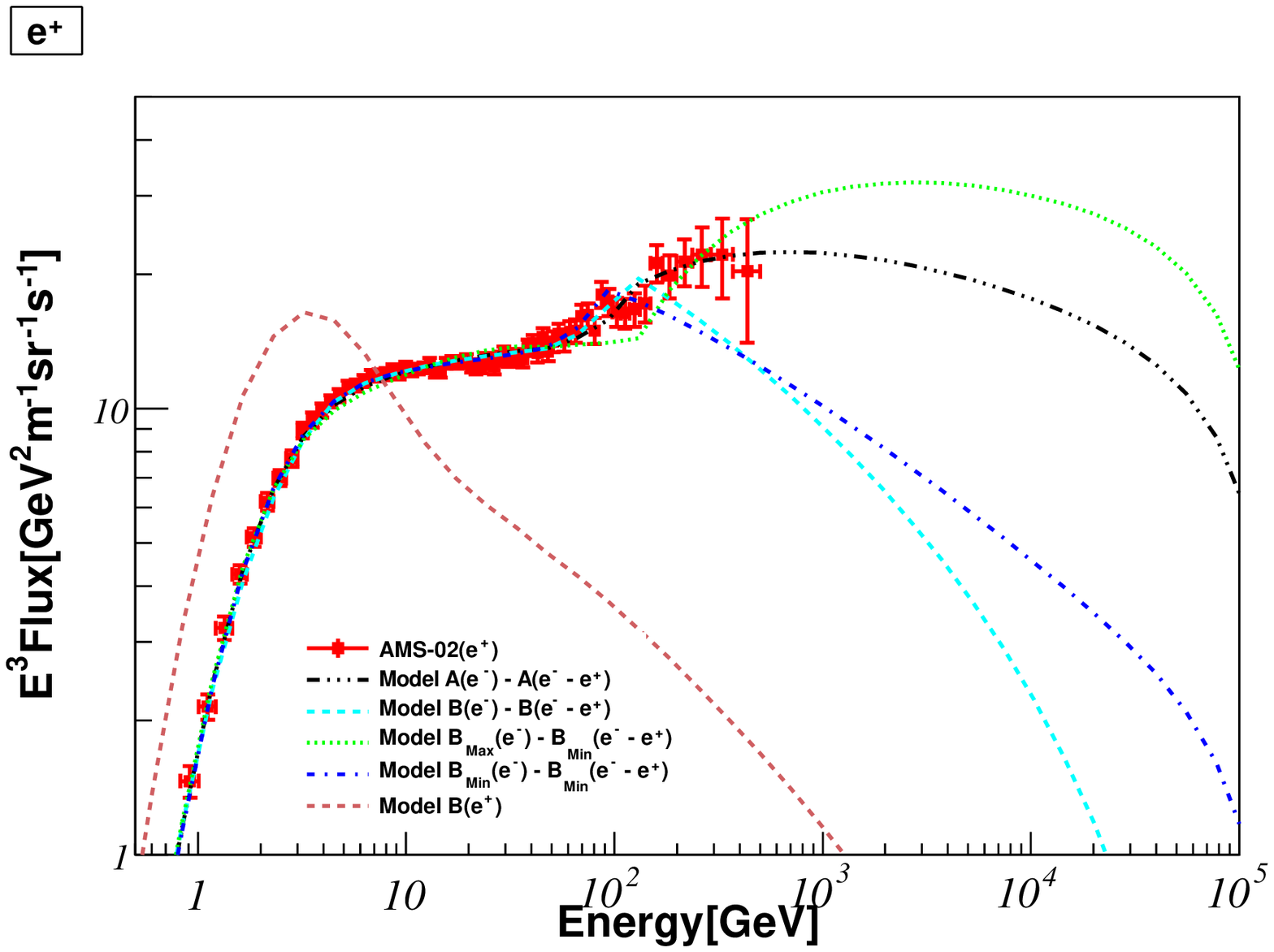}\includegraphics[width=0.49\textwidth]{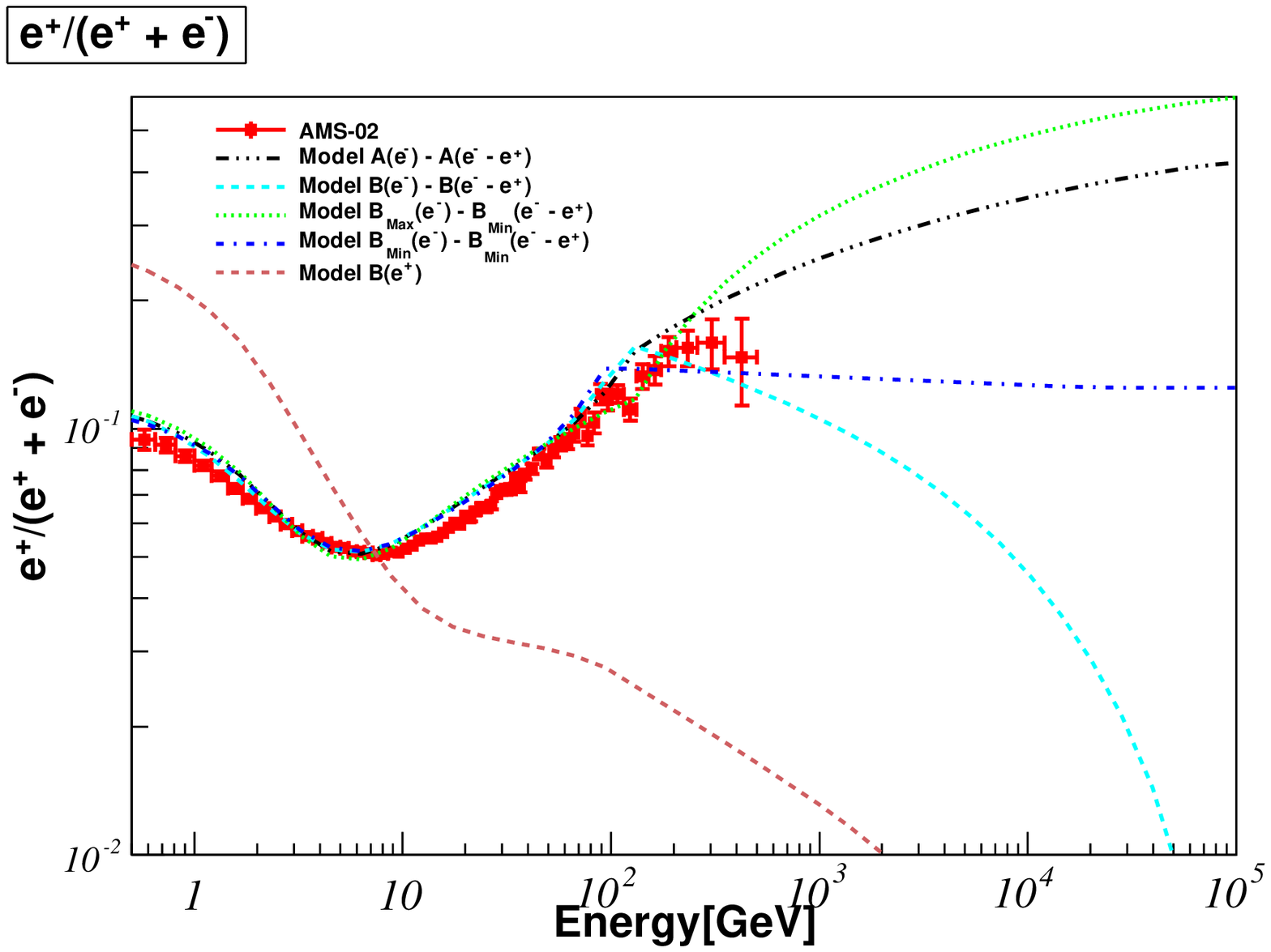}
\caption{(Left) CR positron flux and (right) positron fraction derived from the differences  between CR electrons of $e^{-}$ cases and the primary electrons of $e^{-}$-$e^{+}$ case in model A and B. 
In the lines, some are the differences of the best-fit fluxes, others are relevant to the flux bounds from Table \ref{tab:para_MinMax}. 
The flux of CR positrons and positron fraction from the astrophysical background are calculated from the $e^{-}$ case in model B, whose notation is $e^{+}$ in the brackets. $e^{-}$ and $e^{-}$-$e^{+}$ in the brackets denote $e^{-}$ and $e^{-}$-$e^{+}$ case, respectively.
The flux of CR positrons\cite{Aguilar:2014mma} and positron fraction\cite{Accardo:2014lma} from AMS-02 are also drawn.}
 \label{fig:positronPre}
\end{figure}

Based on the best-fits and bounds of CR electron and primary electron fluxes, CR positron spectrum may be predicted from their differences between  the $e^{-}$ and $e^{-}$-$e^{+}$ cases in model A-B.  
In the Figure \ref{fig:positronPre}, The predicted CR positron flux and positron fraction are drawn with the color lines.
In these lines, some are derived from the differences between best-fit fluxes of CR electrons of $e^{-}$ cases and the primary electrons of $e^{-}$-$e^{+}$ case in model A-B. The relevant best-fit parameters are shown in Table \ref{tab:para_electronMinusPositon} and \ref{tab:para_electron}. 
The other lines are from the differences between maximal and minimal fluxes of $e^{-}$ case and minimal flux of $e^{-}$-$e^{+}$ case in model B from Table \ref{tab:para_MinMax}. 
As seen in Figure \ref{fig:positronPre}, the rapidly damping spectrum to be interpreted as dark matter\cite{Lin:2014vja} is shown, which is relevant to the difference between best-fit fluxes of $e^{-}$ and $e^{-}$-$e^{+}$ case in model B. And the other predicted spectra interpreted as pulsar wind nebulae (PWN) are found in the paper\cite{DiMauro:2014iia}.

\begin{table}[htb]
\begin{center}
\begin{tabular}{lrrrr}
\hline\hline
Para.&$B^{e^{-} - e^{+}}_{\scriptsize\mbox{Min}}$&$B^{e^{-} - e^{+}}_{\scriptsize\mbox{Max}}$&$B^{e^{-}}_{\scriptsize\mbox{Min}}$&$B^{e^{-}}_{\scriptsize\mbox{Max}}$\\
\hline
$N_e$&1.2856 &1.2893 &1.3573 &1.3616 \\
$\gamma^{e}_{1}$&1.6067 &1.6104 &1.6114 &1.6141 \\
$\gamma^{e}_{2}$&2.7517 &2.7456 &2.7341 &2.7277 \\
$\rho^{e}_{br2}$&0.0930 &0.1786 &0.0685 &0.1142 \\
$\gamma^{e}_{3}$&2.5350 &2.3450 &2.5350 &2.4080 \\
\hline
$Z_h$&2.9527 &2.9525 &2.9545 &2.9547 \\
$D_0$&1.7534 &1.7499 &1.7531 &1.7481 \\
$\delta$&0.2954 &0.2961 &0.2954 &0.2964 \\
$v_{Alfven}$&38.647 &38.620 &38.639 &38.583 \\
\hline
$N_p$&4.5226 &4.5226 &4.5226 &4.5228 \\
$\gamma^{p}_{1}$&1.7796 &1.7784 &1.7795 &1.7781 \\
$\gamma^{p}_{2}$&2.4678 &2.4671 &2.4678 &2.4670 \\
\hline\hline
\end{tabular}
\end{center}
\caption{
The bounds of propagation parameters from the $e^{-}$-$e^{+}$ and $e^{-}$ cases of model B  at 99\% C.L.. $B^{e^{-} - e^{+}}_{\scriptsize\mbox{Min}}$ and  $B^{e^{-} - e^{+}}_{\scriptsize\mbox{Max}}$ represent respectively the minimal and maximal flux of primary electrons in the $e^{-}$-$e^{+}$ case of model B.  $B^{e^{-}}_{\scriptsize\mbox{Min}}$ and  $B^{e^{-}}_{\scriptsize\mbox{Max}}$ are for the flux of CR electrons of $e^{-}$ case. 
}
\label{tab:para_MinMax}
\end{table}
\begin{figure}
\includegraphics[width=0.49\textwidth]{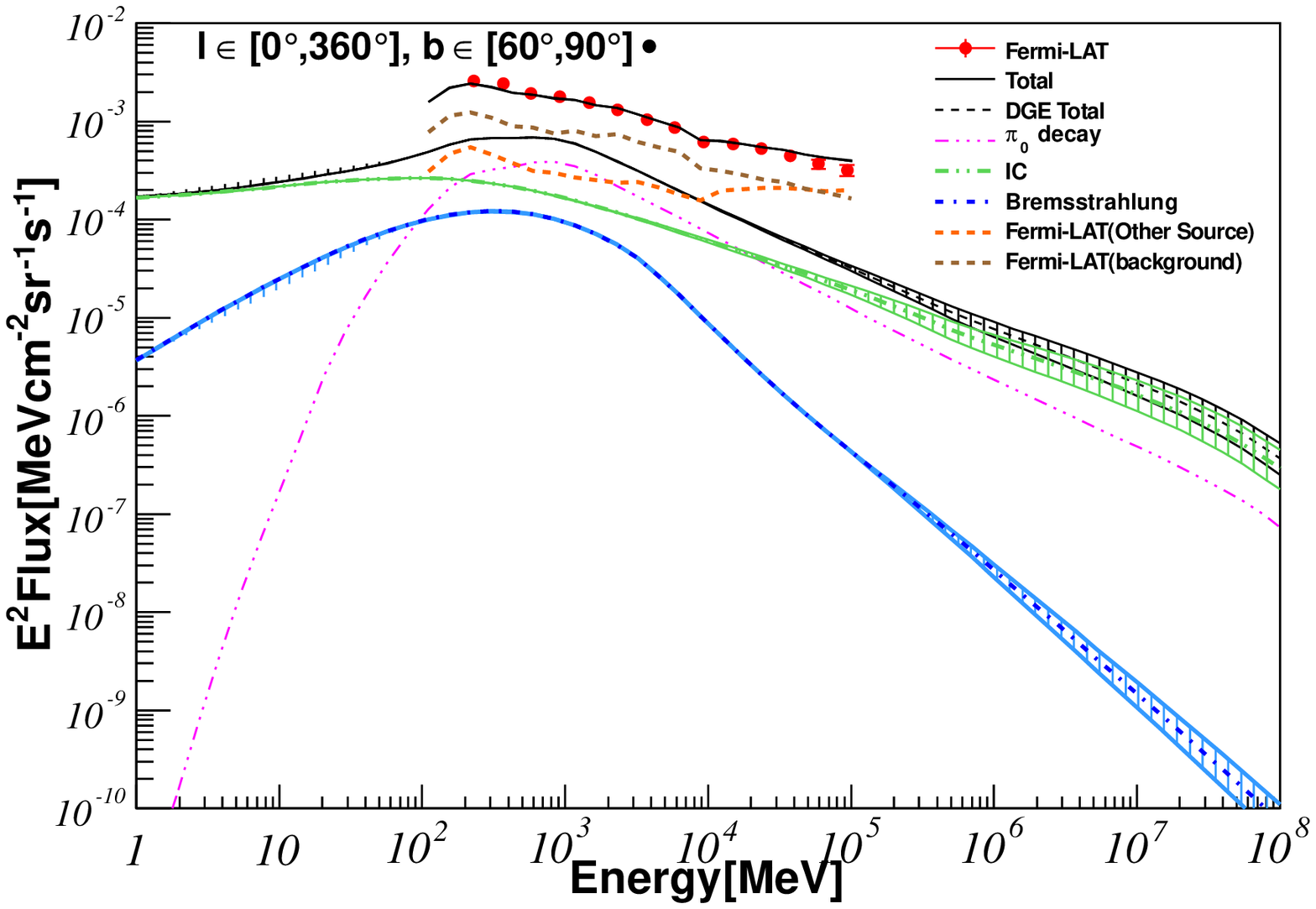}\includegraphics[width=0.49\textwidth]{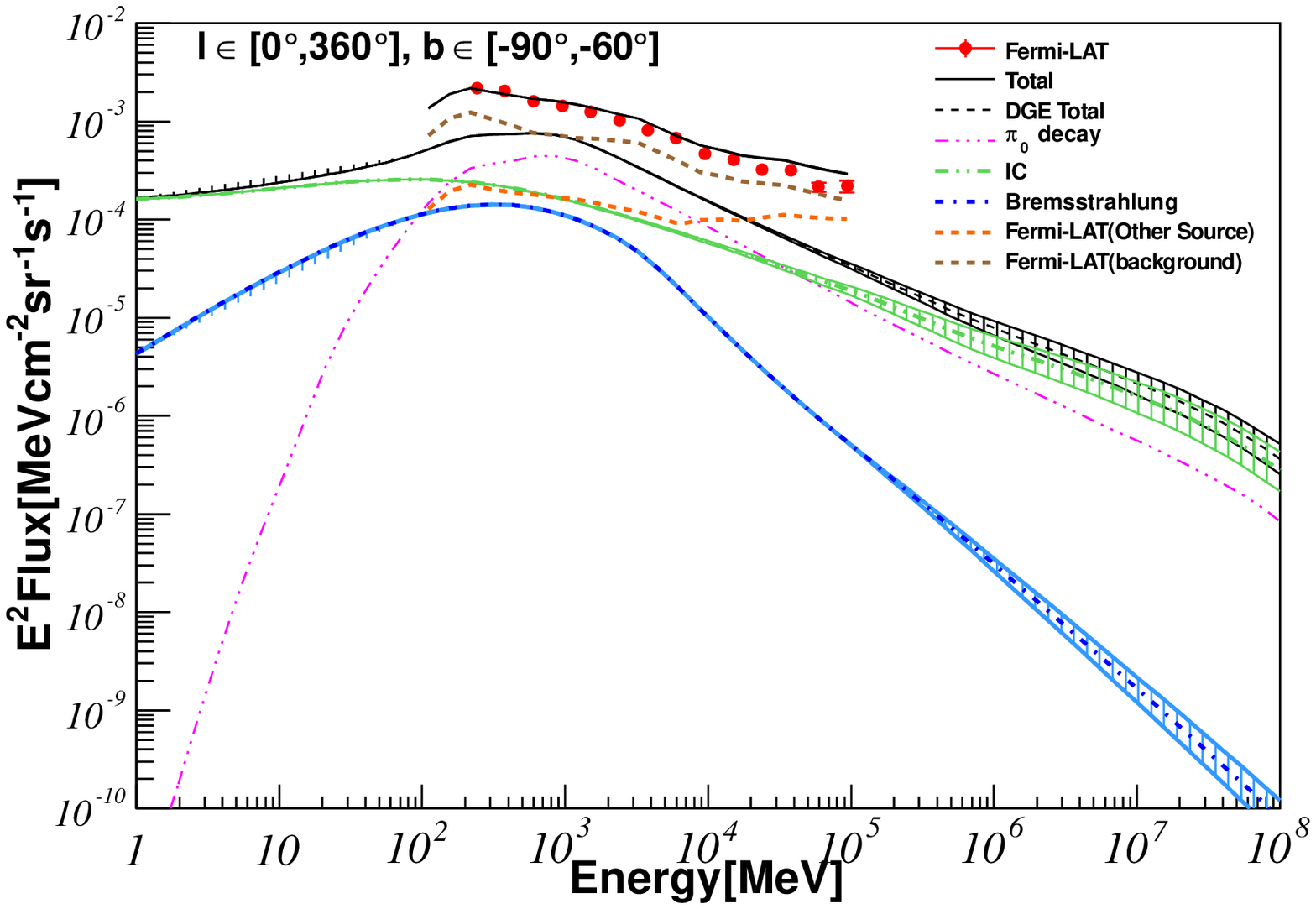}
\includegraphics[width=0.49\textwidth]{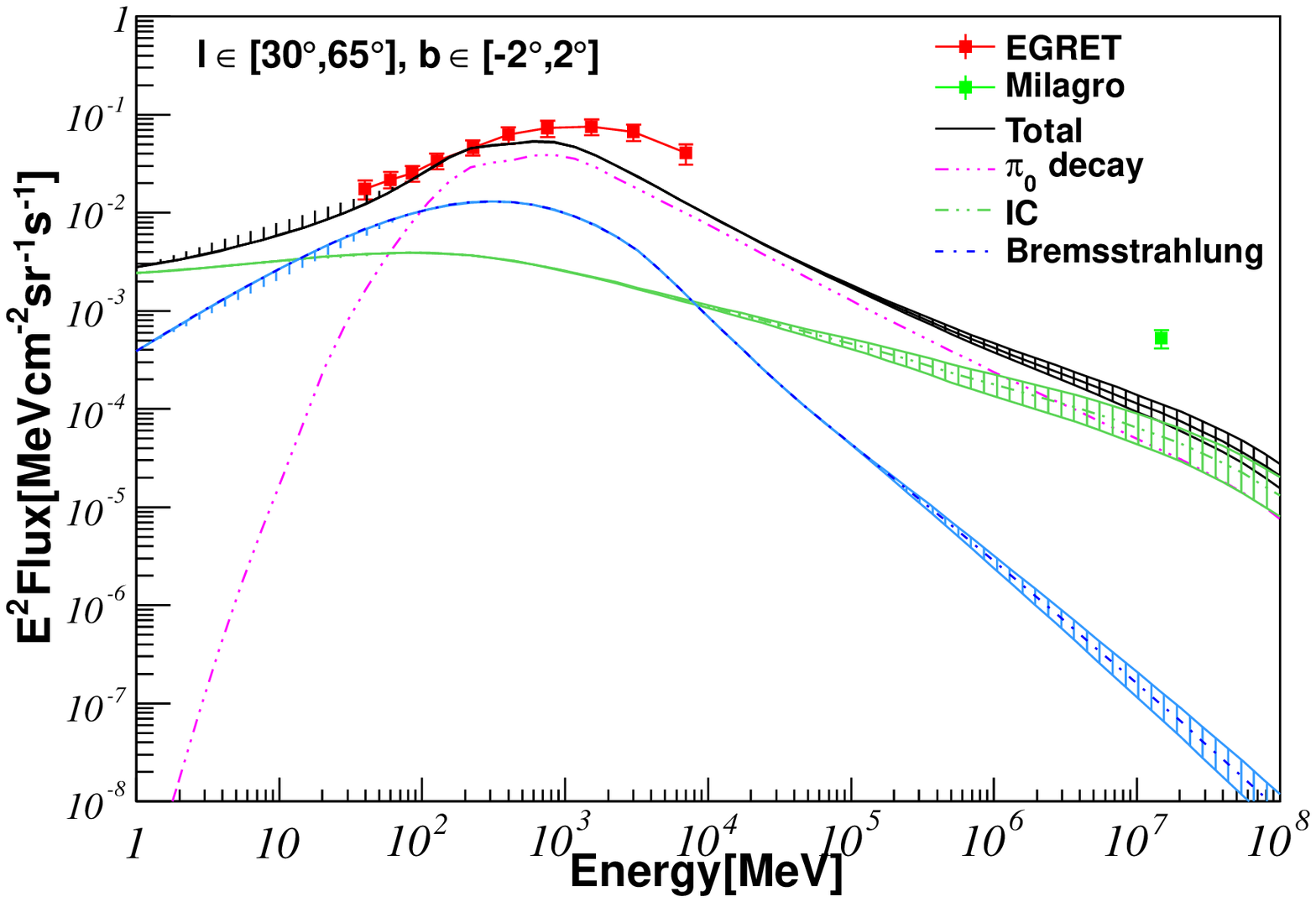}\includegraphics[width=0.49\textwidth]{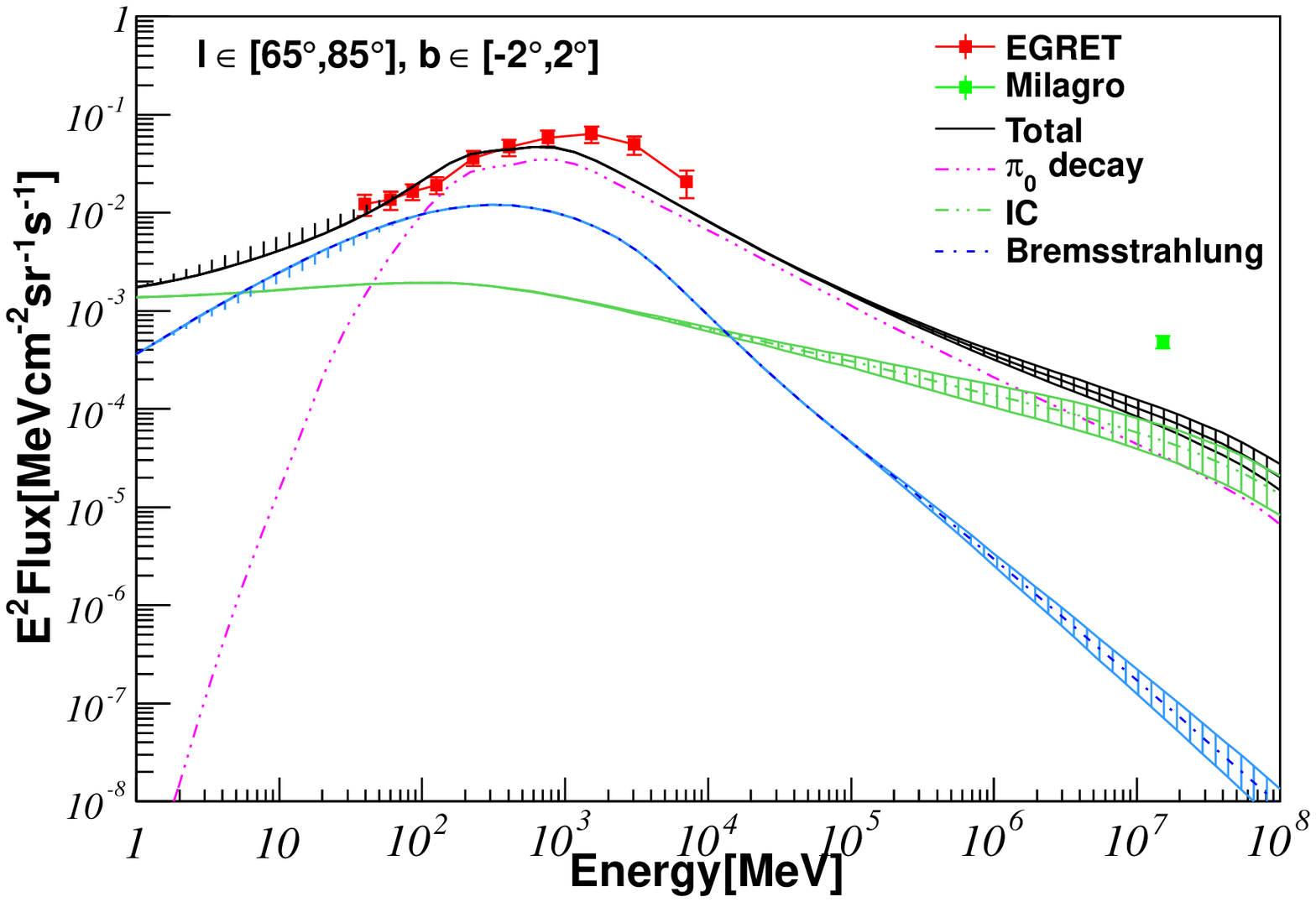}
\caption{(First row) the flux of diffuse gamma rays from all sky($l\in[0^\circ,360^\circ]$) of the high latitude Galaxy (left) ($b\in[60^\circ,90^\circ]$) and (right) ($b\in[-90^\circ,-60^\circ]$). 
The gamma ray fluxes of the total, background and the other sources from Fermi-LAT\cite{FermiLAT:2012aa} are also drawn. 
In the legends, DGE total denotes the total components of diffuse gamma rays and Total denotes total gamma rays including the diffuse gamma rays, other sources and background.
(Second row) the flux of diffuse gamma rays from (left) the inner Galaxy ($l\in[30^\circ,65^\circ]$) and (right) the Cygnus region($l\in[65^\circ,85^\circ]$) in the range of Galactic latitude $b\in[-2^\circ,2^\circ]$. 
The flux of diffuse gamma rays from Milagro\cite{Abdo:2008if} and EGRET\cite{Hunger:1997we} are also drawn. 
In the legends, \textbf{Total} denotes the total components of diffuse gamma rays. 
The filling regions are drawn from the maximal and minimal fluxes of diffuse gamma rays calculated on the parameters of Table \ref{tab:para_MinMax}.
These best-fit fluxes are calculated by use of the parameters of $e^{-}$ case in model C from the Table \ref{tab:para_electron}. 
As the $\pi^0$-decay component of diffuse gamma rays have the less uncertainties than 2\%, the relevant regions are the same as the width of best-fit lines and not filled. 
}
 \label{fig:MilagroSky}
\end{figure}
\subsection{The spectra of the diffuse gamma rays and its comparison with CRs}
In the $e^{-}$ cases of model C, the diffuse gamma rays of two regions from Milagro experiments do not constrain indirectly the more high energy spectrum of CR electrons.
As seen in Table \ref{tab:chisquares_electron}, the best-fit  $\chi^2$ relevant to the AMS-02 electron of model C is almost the same as the model B. 
In Figure \ref{fig:AMS02_ProtonAndBC}, CR electrons, CR protons and B/C are not distinctly discriminated between the model B-C.
These situations indicate that the CR electrons including the positron excess do not derive the more flux of IC component of diffuse gamma rays to fit two regions of Milagro.
In order to analyze the predicted fluxes of the diffuse gamma rays from the uncertainties of CR electrons in the context, 
the bounds of the diffuse gamma rays are calculated on the parameters of $e^{-}$ case in model B from Table \ref{tab:para_MinMax} by GALPROP package.

In Figure \ref{fig:MilagroSky}, the fluxes of three components of diffuse gamma rays from the some regions are shown. 
The filling regions are drawn with the maximal and minimal fluxes of diffuse gamma rays  at 99\% C.L. from Table \ref{tab:para_MinMax}.
As the $\pi^0$-decay component of diffuse gamma rays have uncertainties less than 2\%, the relevant regions are invisible and not shown.
In the first row of Figure \ref{fig:MilagroSky}, Fermi-LAT data inlcudes the gamma rays of the diffuse, background and point sources in the hight latitudes. As seen, below TeV, the total diffuse gamma rays predicted by $e^{-}$ case in model B  are consistent with Fermi-LAT data. In the energy ranges of Fermi-LAT data, the $\pi^0$-decay are dominant in the three components of  gamma rays.
As the flux of CR protons has little uncertainties, the uncertainties of diffuse gamma rays is mainly from IC component predicted by CR electrons, which are represented by the filling regions of Figure \ref{fig:MilagroSky}.

As seen in the second row of Figure \ref{fig:MilagroSky}, by comparison to the data of Milagro, the flux of diffuse gamma rays, derived from CR protons and electrons favored by AMS-02 and HESS, is weaker and not used to interpret gamma rays excess of Milagro, though the relevant flux of positron excess is well extended by the data of AMS-02 and HESS. It implies that the interpretation of positron excess can not agree spontaneously with the gamma rays excess. 

\begin{figure}
\includegraphics[width=0.49\textwidth]{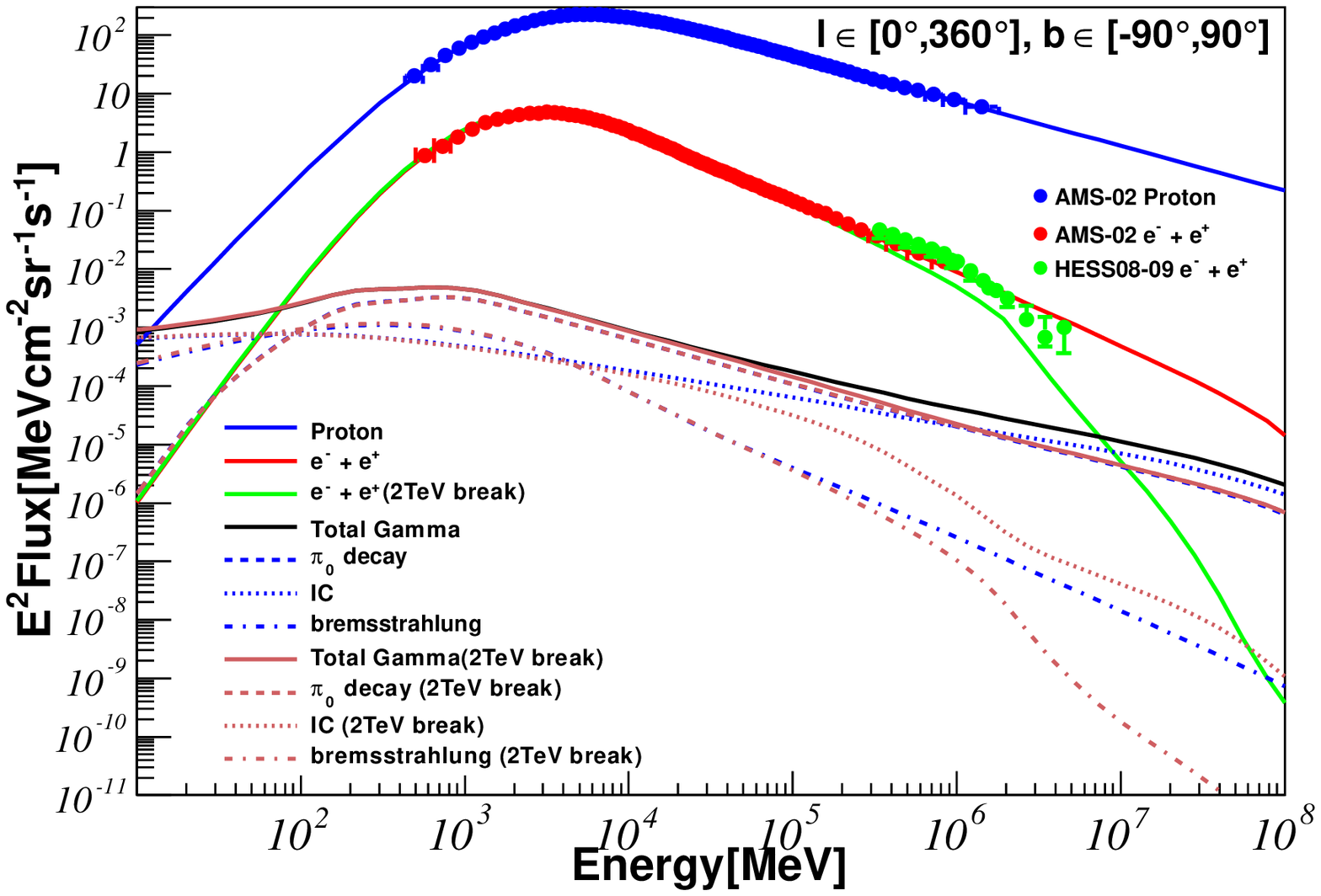}\includegraphics[width=0.49\textwidth]{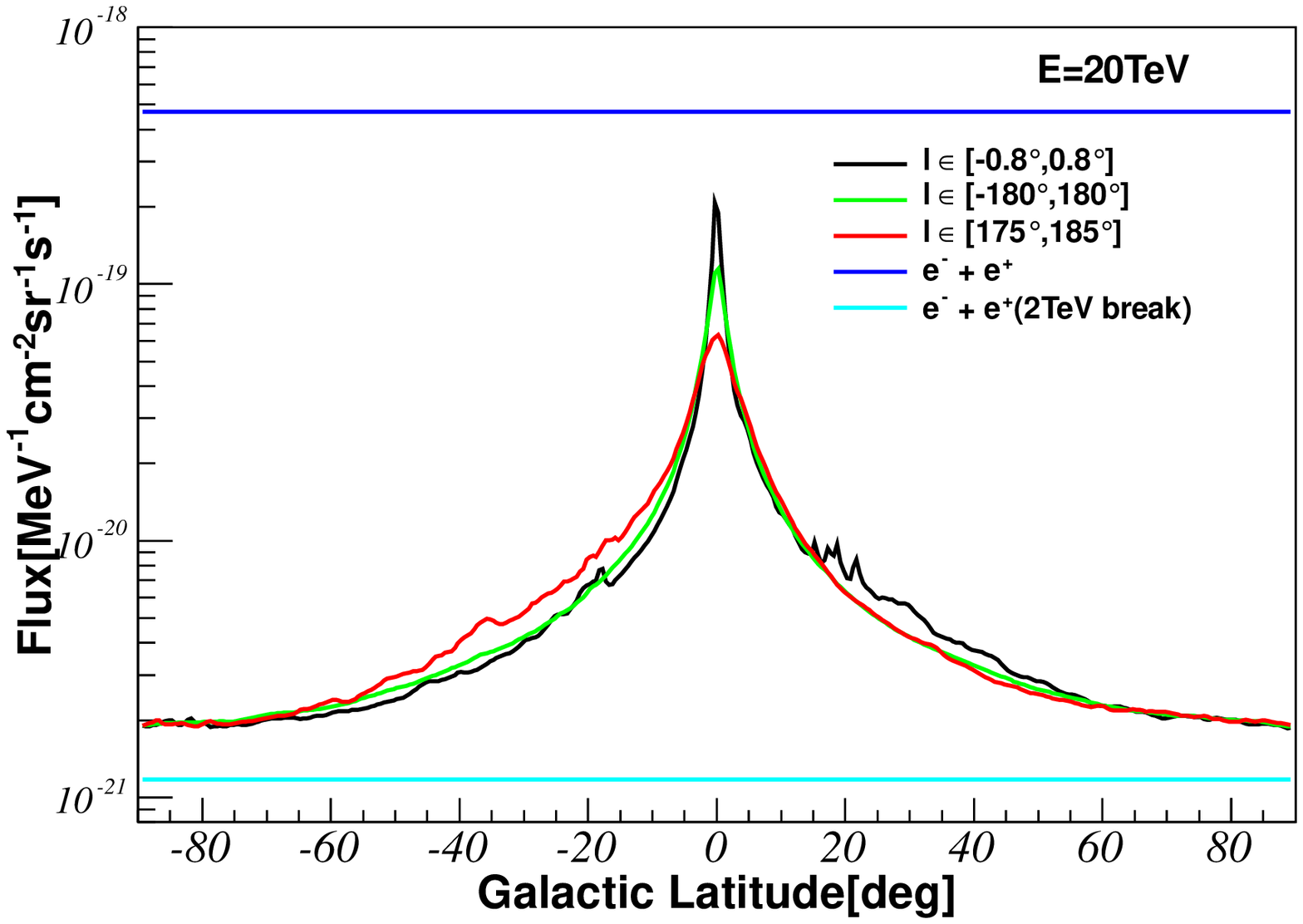}
\caption{
(Left)comparisons between the fluxes of diffuse gamma rays, CR electrons and CR protons. The components of diffuse gamma rays are drawn.  
The fluxes of CR protons, $\pi^0$-decay gamma rays or the other of no brackets are relevant to the best-fit $\chi^2$ of $e^{-}$ case in model C.  2 TeV break in the brackets means the fluxes are relevant to the  best-fit $\chi^2$ of $e^{-}$ case in model B${}_{\scriptsize\mbox{2TV}}$.  CR electrons from AMS-02\cite{Aguilar:2014mma,Aguilar:2014fea} and HESS\cite{Aharonian:2008aa,Aharonian:2009ah} experiments and CR protons from AMS-02\cite{Aguilar:2015ooa} are also drawn. 
 (Right) comparison betweens the 20 TeV flux of  the CR electrons and the diffuse gamma rays at the Galactic latitude coordinate. The fluxes of no brackets and 2 TeV break in the brackets are relevant to the same models as the left figure. The relevant parameters are from the Table \ref{tab:para_electron}.}
 \label{fig:beComp}
\end{figure}

In the comparisons between the components of diffuse gamma rays, it is clear that the $\pi^0$-decay gamma rays are dominant  in the order of  GeV. 
From the high latitude regions to the low ones, the flux of IC component is dominant in the order of hundred GeV to tens of TeV. 
This is also seen in the left of Figure \ref{fig:beComp}. The fluxes of IC and $\pi^0$-decay components of diffuse gamma rays overlap at the middle energy regions at the high and low latitude, except for the 2 TeV break case.
These results are derived from the fact that the distribution of ISM increases from the high latitude regions to the low ones and the fluxes of ISRF almost have the same intensity out of the core regions of Galaxy. 

In the left of Figure \ref{fig:beComp}, CR electrons and protons from $e^{-}$ case in model C and model B${}_{\scriptsize\mbox{2TV}}$ are shown together. And the fluxes of all-sky diffuse gamma rays derived from them are also drawn. 
From an overall comparison, the flux of CR protons is almost 2-5 orders of magnitude greater than CR electrons and 4-7 orders higher than diffuse gamma rays. 
It means that in the indirect measurements of CR electrons and diffuse gamma rays, the main background is CR protons. 
In details, for the $e^{-}$ case of model B${}_{\scriptsize\mbox{2TV}}$, above 10 TeV the flux of diffuse gamma rays are slightly greater than CR electrons. 
And above 0.1 GeV, the flux of $\pi^0$-decay component of  diffuse gamma rays are completely beyond IC. $\pi^0$-decay are dominant in the three components of  gamma rays.

In the right of Figure \ref{fig:beComp}, the diffuse gamma rays of some typical regions derived from CR protons and electrons relevant to the best-fit $\chi^2$ of $e^{-}$ case in model C are shown at the Galactic latitude coordinate. 
The 20 TeV fluxes of CR electrons relevant to the best-fit $\chi^2$ of $e^{-}$ case in model C and model B${}_{\scriptsize\mbox{2TV}}$ are drawn too. 
It is clear that in the right of Figure \ref{fig:beComp}, if TeV extension of CR electrons has a 2 TeV break, at the regions of $b\in[-90^\circ,-40^\circ]$ and $[40^\circ,90^\circ]$ the flux of diffuse gamma rays will be near the flux of 20 TeV CR electrons, but at the regions of $b\in[-40^\circ,40^\circ]$ the flux of diffuse gamma rays is apparently greater than CR electrons from the high to the low latitude. 
Nevertheless, for the $e^{-}$ case in model C, the flux of diffuse gamma rays is even less than 20 TeV CR electrons and the maximal difference is almost beyond the two orders. 
The minimal difference of that is only at the very small latitudes of the center of Milk Way, where the flux of diffuse gamma rays is close to CR electrons.
From these comparisons, it is known that  if above TeV the spectral features of CR electrons constrained by AMS-02 and HESS are consistent with the experimental measurements of the future, in the heigh latitudes the fluxes of diffuse gamma rays will be very weaker than CR electrons and ignored in the background subtractions of indirectly measured CR electrons based on the air-shower array.   

\section{Conclusions}\label{sec:conclusion}
Based on the conventional diffusion model of CRs, we perform a global analysis of the spectral features of CR electrons and the diffuse gamma rays with the data of AMS-02 and HESS by GALPROP package.
The results show that the spectrum structure of the primary component of CR electrons is not described by a simple power law and the relevant break is around the hundred GeV.
In details, we do some analysis of confidence intervals to illustrate the changes of the second reference rigidity and the two indices of primary electrons. 
At the 99\% C.L.,  based on the constraint of AMS-02 data alone, the minimal difference between the second and third indices at the reference 50GV is only 0.075, but the maximal difference does not converge reasonably, which means that AMS-02 data does not exclude primary electron spectrum from a simple power law.  With the constraint of AMS-02 and HESS data, the third index decreases from 2.54 to 2.35 at 99\% C.L., but the second one is only in the range 2.746 - 2.751.  Apparently, the spectrum of primary electrons favors the feature distinguished from a simple power law. 

Above TeV, the often used 2 TeV break case are also analyzed. The result show that the precise AMS-02 data alone do not distinguish the 2 TeV break and a simple power law. In combination of AMS-02 and HESS data, we find that the spectrum of CR electrons do not need the TeV breaks and favor a power law above the hundred GeV.

With the difference between the CR electrons and primary electrons constrained by AMS-02 and HESS, the bounds of TeV extensions of positron excess are predicted. 
In the bounds, some spectra are damping rapidly with the energy, which is consistent with the features of dark matter annihilation and decay\cite{Lin:2014vja}, and the other are similar to the spectra of PWN\cite{DiMauro:2014iia}.

Galactic diffuse gamma ray emission is mainly from the interactions of cosmic rays with the interstellar medium of the Milky Way and involve $\pi^0$-decay, IC scattering and bremsstrahlung.
Based on the data of HESS and AMS-02 and Milagro, we also perform an analysis on the origins of Galactic diffuse gamma rays. 
Below TeV, the total diffuse gamma rays predicted by HESS and AMS-02 are consistent with Fermi-LAT data.
Above TeV, the flux of IC component, derived from the up-limit of CR electrons at 99\%C.L., does not  fit to the Milagro data.
As the up-limit fluxes of CR electrons include the same flux as positron excess, the interpretation of positron excess can not agree spontaneously with the diffuse gamma rays excess from Milagro data.
It also implies that except for dark matter or pulsar, the other sources of diffuse gamma rays  contribute to the measurement data of Milagro.

In the analysis of the identification of Galactic diffuse gamma rays, inverse Compton scattering(IC) is the dominant component in the range of the hundred GeV to tens of TeV respectively from the high latitude regions to the low ones. 

In this paper, the differences between the TeV fluxes of CR electrons and diffuse gamma rays are also analyzed.
The results show that the TeV breaks of CR electron spectrum determine whether the 20 TeV flux of diffuse gamma rays is greater or less than CR electrons.
The TeV extension of CR electrons favored by AMS-02 and HESS has no breaks of TeV and the relevant flux is greater than the diffuse gamma rays in most regions of Milk Way. 
Inversely, if TeV extension of CR electrons has a strong TeV break, the 20 TeV flux of diffuse gamma rays are greater than CR electrons at the regions of $b\in[-40^\circ,40^\circ]$. Out of these regions, the flux of diffuse gamma rays is near the CR electrons.  

As above GeV the flux of CR electrons, constrained by AMS-02 and HESS is much greater than the diffuse gamma rays at high latitudes, background subtractions do not consider the diffuse gamma rays in the indirectly measured CR electrons based on the air-shower array. For the measurements of diffuse gamma rays, CR electrons are not ignorable component in the  subtraction background.

\textit{Note added:}
At the same time when the fitting schemes are confirmed by the authors of this paper, Yu-Feng Zhou also proposes the $e^{-}$-$e^{+}$ case of model A as the primary electron background in the research of dark matter concerned.  The relevant works will show the similar fitting data, which are consistent with this paper.
When we finalize the first version of the manuscript, a preprint with similar analysis\cite{Li:2014csu} has been uploaded to arXiv near the submit date. 
The paper \cite{Li:2014csu} mainly focus the excess components of CR electrons with Markov chain Monte Carlo (MCMC) and indicate the similar conclusions about the primary electrons.  
\section*{Acknowledgments}
We thank Yue-Liang Wu and Yu-Feng Zhou for the helpful discussions on the uncertaines of CR electron spectrum and the interpretation of positron excess. 
The numerical calculations were done  by use of the Cluster named laohu of NAOC.
We thank the Cluster administrators: Chang-Hua Li, Chen-Zhou Cui, etc. for the help concerned.
We thank the referee for the improvement of this paper. Some conclusions are derived from the referee's objections and comments. 
This work is supported by the Knowledge Innovation Programs of the Chinese Academy of Sciences.

\clearpage
\bibliography{../../Database/library,../../Database/gammaRayConcerned,../../Database/AMS02concerned,../../Database/CosmicRaysExperiments,../../Database/analysisPaperOfCRs}
\bibliographystyle{JHEP}

\end{document}